\documentclass[twocolumn]{aastex631}
\usepackage{amsmath}
\usepackage[]{overpic}
\usepackage{multirow}
\usepackage{booktabs}
\usepackage{array, multirow, bigdelim, makecell, booktabs}

\newcommand\nSample{16,791 }
\newcommand\nHECATE{117 }
\newcommand\nAsiago{2,563 }
\newcommand\nNearby{2,680 }
\newcommand\nALeRCE{14,111 }
\newcommand\ntest{4,784 }

\newcommand\IDIA{\affiliation{Data and Artificial Intelligence Initiative (IDIA), Faculty of Physical and Mathematical Sciences, University of Chile, Chile.}}
\newcommand\MAS{\affiliation{Millennium Institute of Astrophysics (MAS), Nuncio Monse\~nor S\'otero Sanz 100, Providencia, Santiago, Chile}}
\newcommand\CMM{\affiliation{Center for Mathematical Modeling (CMM), Universidad de Chile, Beauchef 851, Santiago 8320000, Chile.}}
\newcommand\DAS{\affiliation{Department of Astronomy, University of Chile, Las Condes, Santiago, Chile.}}
\newcommand\DIE{\affiliation{Department of Electrical Engineering, University of Chile, Santiago, Chile.}}
\newcommand\IAUC{\affiliation{Instituto de Astrof{\'{\i}}sica, Facultad de F{\'{i}}sica, Pontificia Universidad Cat{\'{o}}lica de Chile, Santiago, Chile.}}
\newcommand\CAIUC{\affiliation{Centro de Astroingenier{\'{\i}}a, Facultad de F{\'{i}}sica, Pontificia Universidad Cat{\'{o}}lica de Chile, Santiago, Chile.}} 
\newcommand\UV{\affiliation{Instituto de F\'isica y Astronom\'ia, Facultad de Ciencias, Universidad de Valpara\'iso, Gran Breta\!{n}a 1111, Valpara\'iso, Chile.}}
\newcommand\ESO{\affiliation{European Southern Observatory, Karl-Schwarzschild-Strasse 2, 85748 Garching bei M\"unchen, Germany.}}
\newcommand\UNAB{\affiliation{Instituto de Astrofisica, Facultad de Ciencias Exactas, Universidad Andres Bello, Las Condes, Santiago, Chile.}}
\newcommand\UChicago{\affiliation{Department of Astronomy and Astrophysics, The University of Chicago, Chicago, IL 60637, USA.}}
\newcommand\UCHFisica{\affiliation{Departamento de Física, Facultad de Ciencias F\'isicas y Matem\'aticas, Universidad de Chile, Santiago, Chile.}}
\newcommand\BCNSI{\affiliation{Barcelona School of Informatics, Polytechnic University of Catalonia, 08034 Barcelona, Spain.}}
\newcommand\FisBCN{\affiliation{Faculty of Physics, University of Barcelona, 08028 Barcelona, Spain.}}
\newcommand\IUV{\affiliation{Escuela de Ingeniería Eléctrica, Pontificia Universidad Católica de Valparaíso, Valparaíso, Chile.}}
\newcommand\PhysStanford{\affiliation{Department of Physics, Stanford University, 382 Via Pueblo Mall, Stanford, CA 94305, USA.}}
\newcommand\KavliStanford{\affiliation{Kavli Institute for Particle Astrophysics \& Cosmology, P.O. Box 2450, Stanford University, Stanford, CA 94305, USA.}}
\newcommand\IllinoisAstro{\affiliation{Department of Astronomy, University of Illinois at Urbana-Champaign, 1002 W. Green St., IL 61801, USA.}}
\newcommand\CAS{\affiliation{Center for Astrophysical Surveys, National Center for Supercomputing Applications, Urbana, IL, 61801, USA.}}
\newcommand\NSF{\affiliation{National Science Foundation Graduate Research Fellow.}}
\newcommand\UAustral{\affiliation{Instituto de Informática, Universidad Austral, Valdivia, Chile.}}
\newcommand\Fintual{\affiliation{Fintual Administradora General de Fondos S.A., Santiago, Chile.}}
\newcommand\DO{\affiliation{Data Observatory Foundation, Santiago, Chile.}}
\newcommand\UdeC{\affiliation{Department of Computer Science, University of Concepción, Concepci\'on, Chile.}}

\newcommand\alercelink[1]{\href{https://alerce.online/object/#1}{#1}}

\shorttitle{AASTeX v6.3.1 Sample article}
\shortauthors{F\"orster et al.}
\graphicspath{{./}{figures/}}
\begin{document}

\title{DELIGHT: Deep Learning Identification of Galaxy Hosts of Transients using Multi-resolution Images}

\author[0000-0003-3459-2270]{Francisco F\"orster}
\IDIA\MAS\CMM\DAS
\author[0000-0002-8722-516X]{Alejandra M. Mu\~noz Arancibia}
\MAS\CMM
\author[0000-0003-3627-0216]{Ignacio Reyes}
\CMM\MAS

\author[0000-0003-4906-8447]{Alexander Gagliano}
\IllinoisAstro\CAS\NSF

\author{Dylan Britt}
\PhysStanford\KavliStanford
\author[0000-0001-9532-9906]{Sara Cuellar-Carrillo}
\IUV
\author[0000-0001-9735-6051]{Felipe Figueroa-Tapia}
\UV
\author[0000-0002-5283-933X]{Ava Polzin}
\UChicago
\author{Yara Yousef}
\BCNSI\FisBCN

\author[0000-0002-2045-7134]{Javier Arredondo}
\MAS\CMM
\author{Diego Rodr\'iguez-Mancini}
\DO

\author{Javier Correa-Orellana}
\MAS\IAUC

\author[0000-0001-7868-7031]{Amelia Bayo}
\ESO
\author[0000-0002-8686-8737]{Franz E. Bauer}
\IAUC\CAIUC\MAS
\author[0000-0001-6003-8877]{Márcio Catelan}
\IAUC\CAIUC\MAS
\author[0000-0002-2720-7218]{Guillermo Cabrera-Vives}
\UdeC\MAS
\author[0000-0001-6191-7160]{Raya Dastidar}
\UNAB\MAS
\author[0000-0001-9164-4722]{Pablo A. Est\'evez}
\DIE\MAS
\author[0000-0003-0006-0188]{Giuliano Pignata}
\UNAB\MAS
\author[0000-0002-8606-6961]{Lorena Hernandez-Garcia}
\MAS\UV
\author[0000-0003-3541-1697]{Pablo Huijse}
\UAustral\MAS
\author[0000-0003-3455-9358]{Esteban Reyes}
\Fintual
\author[0000-0003-0820-4692]{Paula S\'anchez-S\'aez}
\ESO\MAS

\author{Mauricio Ramirez}
\UNAB
\author{Daniela Grand\'on}
\UCHFisica
\author[0000-0003-0737-8463]{Jonathan Pineda-García}
\UNAB
\author{Francisca Chabour-Barra}
\DAS\UCHFisica
\author{Javier Silva-Farfán}
\DAS



\begin{abstract}

We present DELIGHT, or Deep Learning Identification of Galaxy Hosts of Transients, a new algorithm designed to automatically and in real-time identify the host galaxies of extragalactic transients. The proposed algorithm receives as input compact, multi-resolution images centered at the position of a transient candidate and outputs two-dimensional offset vectors that connect the transient with the center of its predicted host. The multi-resolution input consists of a set of images with the same number of pixels, but with progressively larger pixel sizes and fields of view. A sample of \nSample galaxies visually identified by the ALeRCE broker team was used to train a convolutional neural network regression model. We show that this method is able to correctly identify both relatively large ($10\arcsec < r < 60\arcsec$) and small ($r \le 10\arcsec$) apparent size host galaxies  using much less information (32 kB) than with a large, single-resolution image (920 kB). The proposed method has fewer catastrophic errors in recovering the position and is more complete and has less contamination ($< 0.86\%$) recovering the cross-matched redshift than other state-of-the-art methods. The more efficient representation provided by multi-resolution input images could allow for the identification of transient  host galaxies in real-time, if adopted in alert streams from new generation of large etendue telescopes such as the Vera C. Rubin Observatory. 

\end{abstract}


\received{Aug 5th, 2022}
\submitjournal{AJ}


\section{Introduction} \label{sec:intro}

A new generation of astronomical surveys is revolutionizing the study of the dynamic sky thanks to their large etendues, i.e., the product between field of view and collecting area, and their massive data processing capabilities. Surveys such as the Zwicky Transient Facility \cite[ZTF,][]{2019PASP..131a8002B, 2019PASP..131g8001G}, {\em Gaia} \citep{2016A&A...595A...1G, 2019IAUS..339...12B}, the Asteroid Terrestrial--Impact Last Alert System, \cite[ATLAS,][]{2018PASP..130f4505T}, and the Legacy Survey of Space and Time \citep[LSST,][]{2009arXiv0912.0201L,2019ApJ...873..111I} are detecting or will detect objects that change their position and/or brightness at rates going from hundreds of thousands to tens of millions of objects per night in almost real-time. These changes are reported in massive alert streams, where alerts contain image cutouts and other relevant information about objects of Solar System, Galactic, or extragalactic origin, but whose nature is not necessarily known in advance. Consequently, a new generation of astronomical alert brokers: the Automatic Learning for the Rapid Classification of Alerts  \cite[ALeRCE,][]{2021AJ....161..242F};  the Alert Management, Photometry and Evaluation of Lightcurves \cite[AMPEL,][]{2019A&A...631A.147N}; the Arizona-NOAO Temporal Analysis and Response to Events System \cite[ANTARES,][]{2018ApJS..236....9N}; Babamul\footnote{\url{https://docs.babamul.dev/}}; Fink \citep{2021MNRAS.501.3272M}; Lasair \citep{2019RNAAS...3...26S}; and Pitt-Google\footnote{\url{https://pitt-broker.readthedocs.io/}} are being developed to provide a fast and reliable filtering and classification to enable the real-time characterization of objects in the stream using the available follow-up resources, including massively multi--plexed spectroscopic instruments such as the 4-metre Multi-Object Spectroscopic Telescope \cite[4MOST,][]{2012SPIE.8446E..0TD}, the Maunakea Spectroscopic Explorer \cite[MSE,][]{2014SPIE.9145E..15S}, the Multi-Object Spectrograph for the VLT \cite[MOONS,][]{2014SPIE.9147E..0NC}, the Subaru Prime Focus Spectrograph \cite[PFS,][]{2014PASJ...66R...1T}, and the Dark Energy Spectroscopic Instrument \cite[DESI,][]{2016arXiv161100036D}.

An important ingredient for the correct classification and characterization of the objects contained in the alert stream is contextual information. In particular, the possible association with a host galaxy in the vicinity of an alert is key for the classification of extragalactic transients. A correct host association can allow the inference of transient properties based on the properties of its host, notably the distance, but also other global and local host properties such as age, stellar mass, star formation activity, and metallicity, that may provide clues about the nature of the transient's progenitor system.

This host galaxy-alert association can be done via crossmatches with a comparison source catalog, selecting for example the nearest normalized distance object \citep{2006ApJ...648..868S, 2016AJ....152..154G}, or by directly studying deep reference images in the same location of the sky \citep{2021ApJ...908..170G}. Both methods have advantages and disadvantages: while catalog association can be very fast, it can lead to catastrophic errors due to the incorrect representation of galaxies with complex geometries; and while direct image association is generally considered to be more reliable, it can require large input images that are slower to retrieve and process than a catalog entry. 

In this work, we propose a method that is both fast and reliable to directly infer the host galaxies of transient candidates from images. It uses a combination of catalog information (source positions and characteristic ellipses); a compact, multi-resolution image at the location of the transient; and a convolutional neural network trained with a sample of \nSample visually selected transient candidate hosts. This method is able to identify the position of the most likely host galaxy using a small amount of input data (32 kB) in a timescale of about 60 milliseconds using a CPU, fast enough to be applied in future alert streams such as that from LSST. Moreover, this method is able to indicate the degree of confidence of the predicted association by using rotation and flip data augmentation in real-time. 

This paper is organized as follows: in Section \ref{sec:methods} we review some of the existing host association methods and give a brief introduction to the proposed method. In Section~\ref{sec:training} we discuss the training set preparation and show some examples. In Section~\ref{sec:delight} we give a detailed description of the proposed method, its architecture and the hyperparameter optimization. In Section~\ref{sec:results} we show the main results of the method, including a comparison with other methods, as well as an analysis of both the errors in angular separation and redshift association. Finally, in Section~\ref{sec:conclusions} we present the conclusions of this work and discuss future steps. 

\section{Host galaxy identification methods} \label{sec:methods}

Historically, the most common method to do host galaxy association has been the visual inspection of images (e.g., the Asiago catalog, \citealt{1999A&AS..139..531B}).  This method is generally reliable and up to now has been seen to outperform other methods in host galaxies with complex geometries, but it does not scale easily to large numbers of transients and neither is it failproof. First, the identification via visual inspection involves some degree of subjectivity that poses challenges for reproducibility. Second, in some instances an image may not be deep enough to detect the host (in the so called \emph{hostless} supernovae), or it may not provide enough information to break association degeneracies between multiple host candidates. In the former case, deeper images or catalogs may be required to get the correct association; and in the latter, spectroscopic data may be necessary to derive the redshift of the transient directly from narrow spectral lines. In this work we focus on those cases where a visual inspection is not possible, because of the large volume of events where the association is needed; and where no additional information is available, because a fast association is required to decide whether to obtain spectroscopic observations of a transient candidate. 

This type of automatic host association can be done in two ways: 1) selecting a source from a reference catalog based on the position of the transient, or 2) using an image centered in the position of the transient to predict a host center and then select a source from a reference catalog based on the predicted position. In both cases, it is possible to train a machine learning algorithm to predict the position given the input, e.g., using a boosted decision tree to select the best source among nearby catalog matches \citep{2020PASP..132h5002S}. We will now discuss some of the methods that are publicly available.

\subsection{Catalog based association methods} \label{sec:cat_methods}

\paragraph{Nearest neighbour (1-NN)} The simplest association method is to find the closest source in projection to the transient in a reference catalog. This has the advantage that no information other than the position of the source is needed, but it can lead to catastrophic errors when the nearest source (in projection) is a star or a smaller, distant galaxy unrelated to the transient. 

\paragraph{Nearest normalized elliptical SExtractor distance (SEx)} An improved version of the previous method is to use normalized angular separations to associate the transient to the source, which has the disadvantage that information about the typical size and orientation of every source is required. The normalized angular separation can be computed as the dimensionless elliptical distance $R$ defined in \cite{2006ApJ...648..868S}, $R^2 = \verb+CXX+ ~ (x_{\rm tr} - x_{\rm gal})^2 + \verb+CXY+ ~ (x_{\rm tr} - x_{\rm gal}) (y_{\rm tr} - y_{\rm gal}) + \verb+CYY+ ~ (y_{\rm tr} - y_{\rm gal})^2$, where $x_{\rm tr}, y_{\rm tr}$ and $x_{\rm gal}, y_{\rm gal}$ are the transient and galaxy positions in pixels, respectively, and where \verb+CXX, CXY+ and \verb+CYY+ are the SExtractor \citep{sextractor} ellipse parameters\footnote{See \url{https://sextractor.readthedocs.io/en/latest/Position.html}}. The main limitation of this method is that galaxies with complex geometries can be broken down into several different ellipses.

\paragraph{Nearest normalized Directional Light Radius (DLR)} As in SEx, this requires information about the typical size and orientation of every source. In this case, a normalized angular separation is inferred as the ratio between the transient to host angular separation and a \emph{Directional Light Radius} \cite[DLR,][]{2016AJ....152..154G, 2018PASP..130f4002S}  computed using the source Stoke parameters as explained in Equations 4 to 9 of \cite{2021ApJ...908..170G}.

\subsection{Image based association methods}

\paragraph{Gradient Ascent method (Grad)} Alternatively, one can infer the position of the host directly from the input images. This is done in the gradient ascent method described in \citet{2021ApJ...908..170G}, that after some pre-processing of PanSTARRS images, including the removal of stellar sources, can infer the position of the host iteratively starting at the position of the transient and gradually moving in the direction of the local surface brightness gradient until a local maximum is reached. 

\paragraph{Deep Learning Identification of Galaxy Hosts of Transients (DELIGHT)} This is the method proposed in this work, that uses as input multi-resolution images of the transient's location, including small and large angular scales, and predicts the position of the host using a convolutional neural network invariant to rotations and flips. The input images are lightweight, taking about 3.5\% of the space that single resolution images of the same region would require.

\section{Data preparation} \label{sec:training}

For the proposed method, DELIGHT, we use a total of \nSample visually selected host galaxies, restricting the sample to galaxies whose candidates are within one arcminute from the host center. This is equivalent to the angular radius of the the Milky Way seen at a distance of 57 Mpc. The sample can be divided into two families:
\begin{enumerate}
    \item \textbf{ALeRCE sample}: a collection of \nALeRCE transient candidate host galaxies visually identified by the ALeRCE broker team between Aug 2019 and May 2022. The transient candidates were first identified in the ZTF public alert stream using the ALeRCE stamp classifier \citep{2021AJ....162..231C} and then  filtered and vetoed with the help of the ALeRCE SN Hunter tool.\footnote{\url{https://snhunter.alerce.online/}} Resulting candidates were used to visually select the most likely host galaxy position among the positions of NASA Extragalactic Database \cite[NED,][]{1991ASSL..171...89H}; Set of Identifications, Measurements and Bibliography for Astronomical Data \cite[SIMBAD,][]{2000A&AS..143....9W}; and Sloan Digital Sky Survey \cite[SDSS,][]{2022ApJS..259...35A} sources superimposed on PanSTARRS DR1 color images using the ipyaladin tool.\footnote{\url{https://github.com/cds-astro/ipyaladin}} This information was reported to the Transient Name Server (TNS\footnote{\url{https://www.wis-tns.org/}}) daily for those candidates that had not been reported based on ZTF data before. 
    \item \textbf{Nearby sample}: a collection of \nAsiago visually identified host galaxies from the Asiago catalog \citep{1999A&AS..139..531B} and \nHECATE host galaxies that have TNS reported SNe in the HECATE catalog \citep{2021MNRAS.506.1896K} of nearby galaxies. Thus, our nearby sample includes \nNearby visually selected host galaxies that are in general closer and with a larger angular size (median SExtractor derived semi-major axis of 14.1\arcsec\  vs 2.4\arcsec) than those obtained in the ALeRCE sample. Note that there is no overlap between the ALeRCE and Nearby samples.
\end{enumerate}

\begin{figure} [ht!]
\centering
 \begin{overpic}[scale=.36,percent]{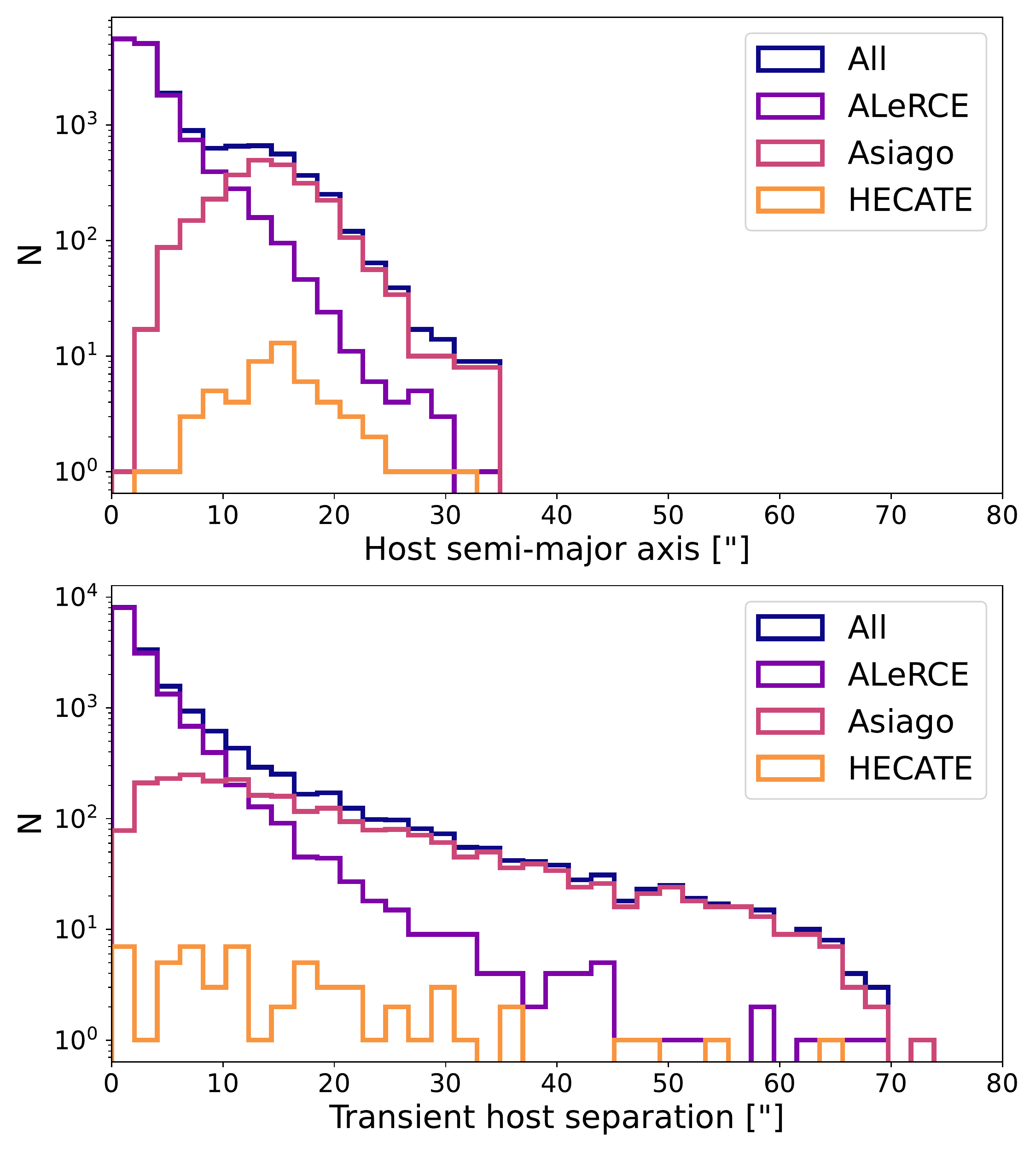}
    \put(17,94) {\large A}
    \put(17,43.5){\large B}
    \end{overpic}
\caption{Angular semi-major axis of transient host galaxies (A) and transient to host angular separation (B) distributions of the input samples. Note that the isophotal semi-major axis is approximately three times the reported semi-major axis \citep{sextractor}. The ALeRCE sample is significantly larger than the nearby sample (\nALeRCE vs \nNearby galaxies, respectively), but it is dominated by smaller and more distant galaxies. Combining both samples helps to reduce the imbalance in host galaxy size (see discussion in Section~\ref{sec:methods}).}
\label{fig:separations}
\end{figure}

The distribution of host galaxy angular semi-major axes and angular separations between the transient and the host galaxy center is shown in Figure~\ref{fig:separations}.  We measured angular semi-major axes using the method explained in Appendix~\ref{sec:appendix_diameter}. Note that the isophotal semi-major axis is approximately three times the SExtractor reported semi-major axis \citep{sextractor}. One can see that the ALeRCE sample is much larger, but that it is dominated by smaller separations since its host galaxies tend to be more distant and smaller in angular size. The Asiago catalog includes transients detected by shallower surveys and in a much larger time span, and the HECATE catalog only contains galaxies within 200 Mpc. By combining both samples we attempt to reduce the imbalance between infrequent transients that have large angular size hosts and more frequent transients that have smaller angular size hosts. 

We further address the imbalance between the sizes of the ALeRCE and nearby samples by selecting a loss function that naturally gives more weight to larger apparent size galaxies, and by generating balanced batches when training the model (see Section~\ref{sec:arch}).

\subsection{Input images} \label{sec:input}

For every transient candidate, we obtain a 120\arcsec$\times$120\arcsec\ (480$\times$480 pixels with 0.25\arcsec/pixel) archival $r$--band PanSTARRS image centered at its location using the \verb+panstamps+ tool.\footnote{\url{https://panstamps.readthedocs.io/en/latest/index.html}} We use only one band in order to ensure a faster image retrieval (it takes about 4 s per image, where each image file size is 920 kB) than would be obtained using multi-band images and we use the $r$--band because it is a better tracer of the bulge, and therefore of the center of  galaxies \citep{2016ApJS..225....6K}. The input images are processed using SExtractor with the SEP tool \citep{2016zndo....159035B} in order to subtract the sky emission and detect significant sources. Pixels that are not within four normalized elliptical radii of any source were masked out (see Section~\ref{sec:cat_methods}).

It is important to note that the catalog positions are not perfect, as many times one can observe a deviation from the true position by several arcseconds and sometimes by tens of arcseconds. Thus, all resulting images were visually examined by a team of 12 people in order to remove incorrectly selected hosts and those where the angular separation between the labeled position and the visually estimated host center was thought to be too large. This step is included when defining the sample of \nSample galaxies.

\setlength{\abovecaptionskip}{-12pt plus 3pt minus 2pt}
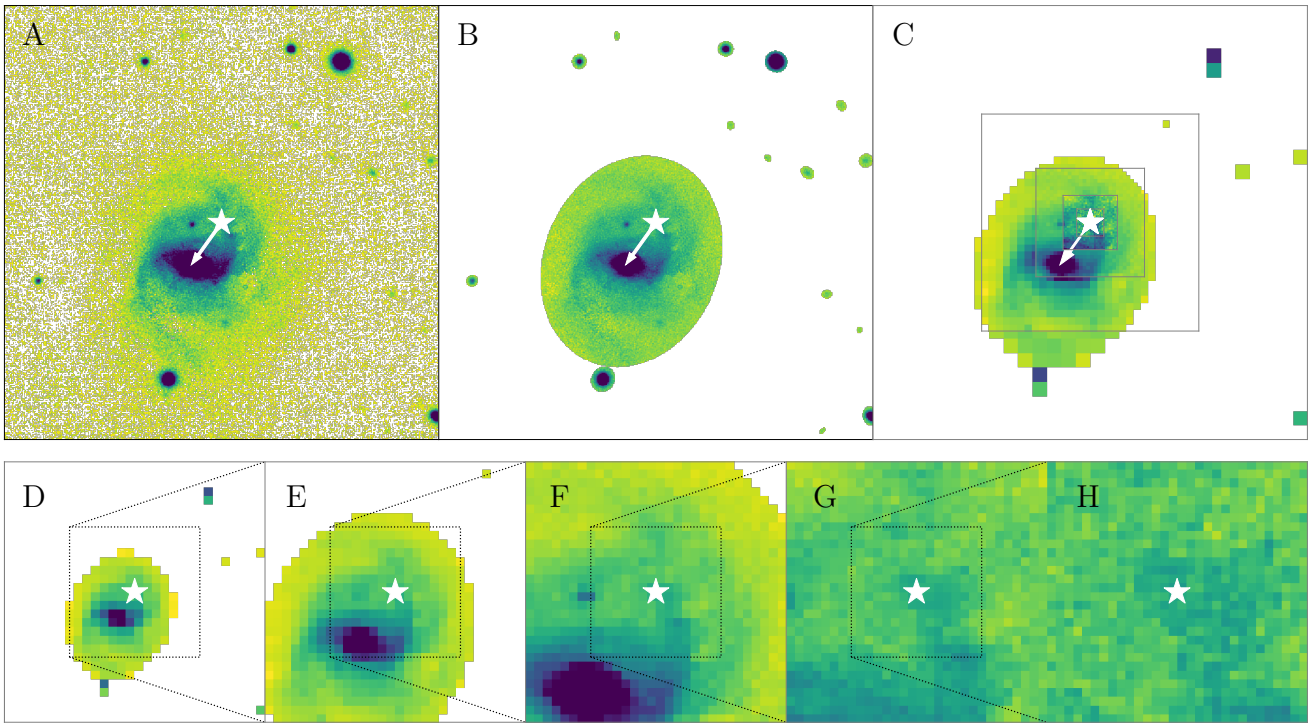
\begin{figure*}[ht!]
  \centering
  \begin{overpic}[scale=.23,percent]{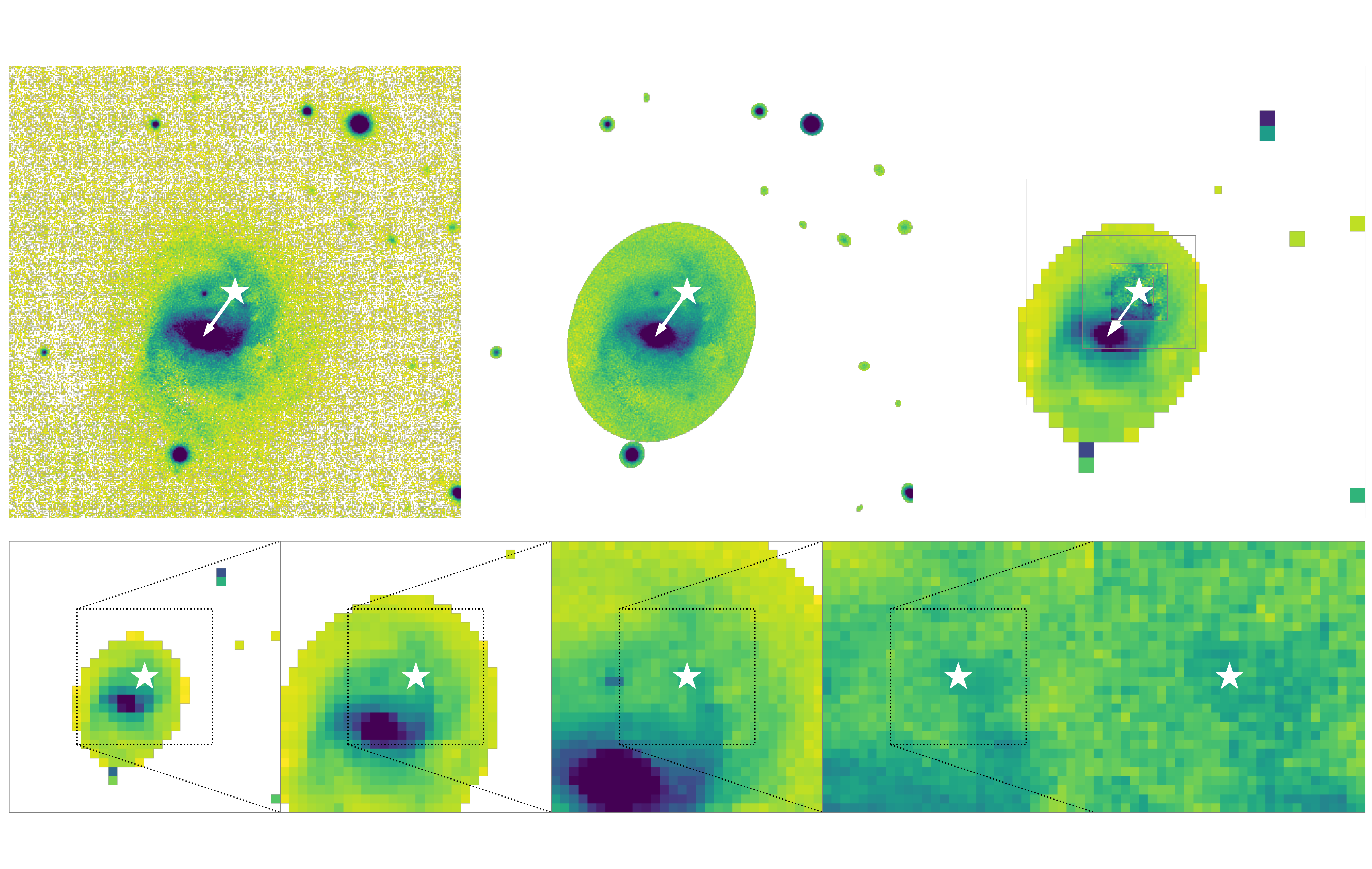}
    \put(2,55.5) {\large A}
    \put(35,55.5){\large B}
    \put(68,55.5){\large C}
    \put(2,20.5) {\large D}
    \put(22,20.5){\large E}
    \put(42,20.5){\large F}
    \put(62,20.5){\large G}
    \put(82,20.5){\large H}
    \end{overpic}

  \caption{    
Training set preparation process: A) initial PanSTARRS $r$-band $480 \times 480$ pixels (0.25\arcsec/pixel) image centered around the position of the transient candidate (white star, in this example SN2004eh), where the output of the model should be the two-dimensional vector connecting the transient and the host galaxy center (white arrow); B) the images are sky subtracted and sources are detected using SExtractor, masking out all pixels that are not within four elliptical dimensionless radii of any source; C) the image is resampled into five images of $30 \times 30$ pixels, all centered around the transient position, but with different fields of view, going from larger to smaller scales as shown in frames D-H. Note that the final multi-resolution image requires 3.5\% of the memory space used by the original image and about 60 milliseconds per source to generate.}
  \label{fig:SN2004eh}
\end{figure*}
\setlength{\abovecaptionskip}{2pt plus 3pt minus 2pt}

The 120\arcsec$\times$120\arcsec, 480$\times$480 pixels, sky subtracted masked images were used to build sets of multi-resolution images made of $n_{\rm levels}$ images. This forces the input to be a more compact representation of the data, enabling its direct inclusion in future enriched alert streams. All $n_{\rm levels}$ images are centered around the position of the candidate and have the same size in pixels, they are squared images with side $480/2^{n_{\rm levels} - 1}$ pixels; but they have different fields of view, with sides 120\arcsec/$2^{i-1}$ with $i \in 1..n_{\rm levels}$ going from larger to smaller scales. Pixels from level $i$ can be built taking the average of 2$\times$2 pixels at the resolution of level $i+1$. All images are normalized between zero and one using the minimum and maximum values between all the levels of a given object. The typical time taken to generate this set is 60 milliseconds per image. An example set of sky subtracted, masked, normalized, five levels multi-resolution images is shown in Figure~\ref{fig:SN2004eh}. In this example, the final image memory size is only 32 kB, much less than the original 920 kB. The detailed algorithm used to construct these images can be found in our public GitHub repository at \url{https://github.com/fforster/delight}. 

\section{Host identification method} \label{sec:delight}

The host identification method is posed as a regression problem where the input data are the multi-resolution images described in Section~\ref{sec:input} and where the outputs are two-dimensional vectors that connect the transients to their hosts in projection. 

The model is based on a convolutional neural network whose architecture is described in Section~\ref{sec:arch}. For the loss function we use the mean squared error between the predicted and labeled offset vectors in units of pixels of the highest resolution image. Using the units of the highest resolution image gives more weight to larger apparent size galaxies, since a small error relative to the apparent size of the galaxy in those cases will result in a much larger relative contribution to the loss function, addressing the imbalance between the host galaxy angular sizes of the ALeRCE and Nearby samples.

In order to make the model invariant to rotations and flips we let the neural network rotate and flip the input images to produce eight different output vectors, that can be averaged after derotating and deflipping to get a final predicted position. Both the dispersion of the previous predictions and the image values at the predicted positions can be used as indicators of when the network may be giving incorrect answers: if the dispersion is too large or if the images at the predicted positions are masked indicates that the predicted position may be incorrect. Both can occur when there are two galaxies that appear equally likely to be the host (see Appendix~\ref{sec:appendix_examples}).

\subsection{Neural network architecture} \label{sec:arch}

The neural network architecture is based on a convolutional  neural network that receives $n_{\rm levels}$ input images of the same size in pixels that are centered around the position of the candidate, as described in Section~\ref{sec:input}. Each set of images is rotated by 90 degrees four times, flipping the image with the fourth rotation, and then further rotating them by 90 degrees three more times. This results in seven more images that are an augmented version of the original set.

\begin{deluxetable*}{ccccccccc}[ht!]
\tablecaption{Summary of the DELIGHT neural network architecture.}
\tablehead{
\colhead{} & \colhead{} & \colhead{} & \colhead{} & \colhead{Type} & \colhead{Description} & \colhead{Activation} & \colhead{Output} & \colhead{\# Parameters}
}
\startdata
 & & & & Input & 5 images of 30$\times$30 pixels &  & 5$\times$30$\times$30 & \\
\multirow{12}{1cm}{Rotations + flips\tablenotemark{a}} & 
\multirow{12}{1em}[17mm]{\ldelim\{{10}{1mm}[]}
& & & Rotation+flip & Rotate and flip input images &  &  8$\times$5$\times$30$\times$30 & \\
\cmidrule{1-9} &  
& \multirow{6}{1cm}{Levels\tablenotemark{b}} &
\multirow{6}{1em}[10mm]{\ldelim\{{6}{1mm}[]} 
& Convolution & 52 channels, 3$\times$3 kernel, no padding & ReLU & 52$\times$28$\times$28 & 520 \\
& & & & Max-pooling & 2$\times$2 max-pooling &  & 52$\times$14$\times$14 & \\
& & & & Convolution & 57 channels, 3$\times$3 kernel, no padding & ReLU & 57$\times$12$\times$12 & 26,733 \\ % 32 4,640
& & & & Max-pooling & 2$\times$2 max-pooling &  & 57$\times$6$\times$6 & \\
& & & & Convolution & 41 channels, 3$\times$3 kernel, no padding & ReLU & 41$\times$4$\times$4 & 21,074 \\ % 32 9,248
& & & & Flatten & Flatten input tensor &  & 656 & \\
\cmidrule{5-9}
& & & & Concatenate & Merge five flattened tensors &  & 3,280 & \\
& & & & Fully connected & 685 neurons & $\tanh$ & 685 & 2,250,766 \\
& & & & Dropout & Regularization of 0.06 at training & & &\\
& & & & Fully connected & 2 neurons &  & 2 & 1,372 \\
\cmidrule{3-9}
& & & & Output & Collect \& undo rotations/flips & & 8$\times$2 &
\enddata
\tablenotetext{a}{Independent processing for eight rotated and flipped sets of five images (levels), sharing weights and biases}
\tablenotetext{b}{Independent processing for five levels, sharing weights and biases.}
\tablecomments{The five input images are first rotated and flipped to produce eight image sets that are processed independently. Within each set, every image is processed independently by a combination of convolutional and max-pooling layers as shown above, that are then flattened, merged and fully connected into two additional layers of 685 and two neurons, respectively. Finally, all outputs are collected and the rotations and flips are undone to produce a set of eight displacement vectors, that can be averaged to produce a single predicted position.}
\label{tab:arch}
\end{deluxetable*}

\setlength{\abovecaptionskip}{-12pt plus 3pt minus 2pt}
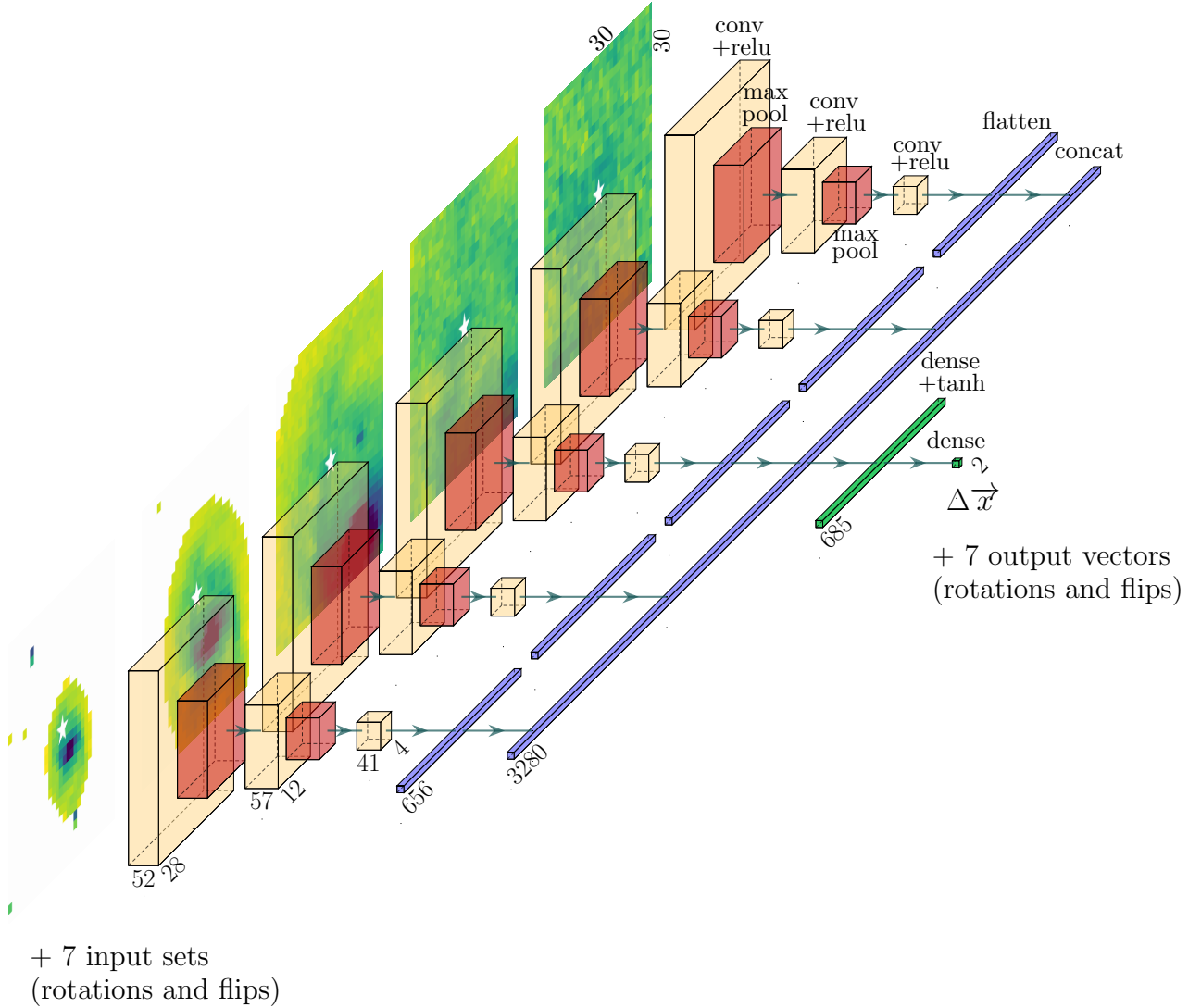
\begin{figure*}[ht!]
  \centering
  \begin{overpic}[scale=.5,percent]{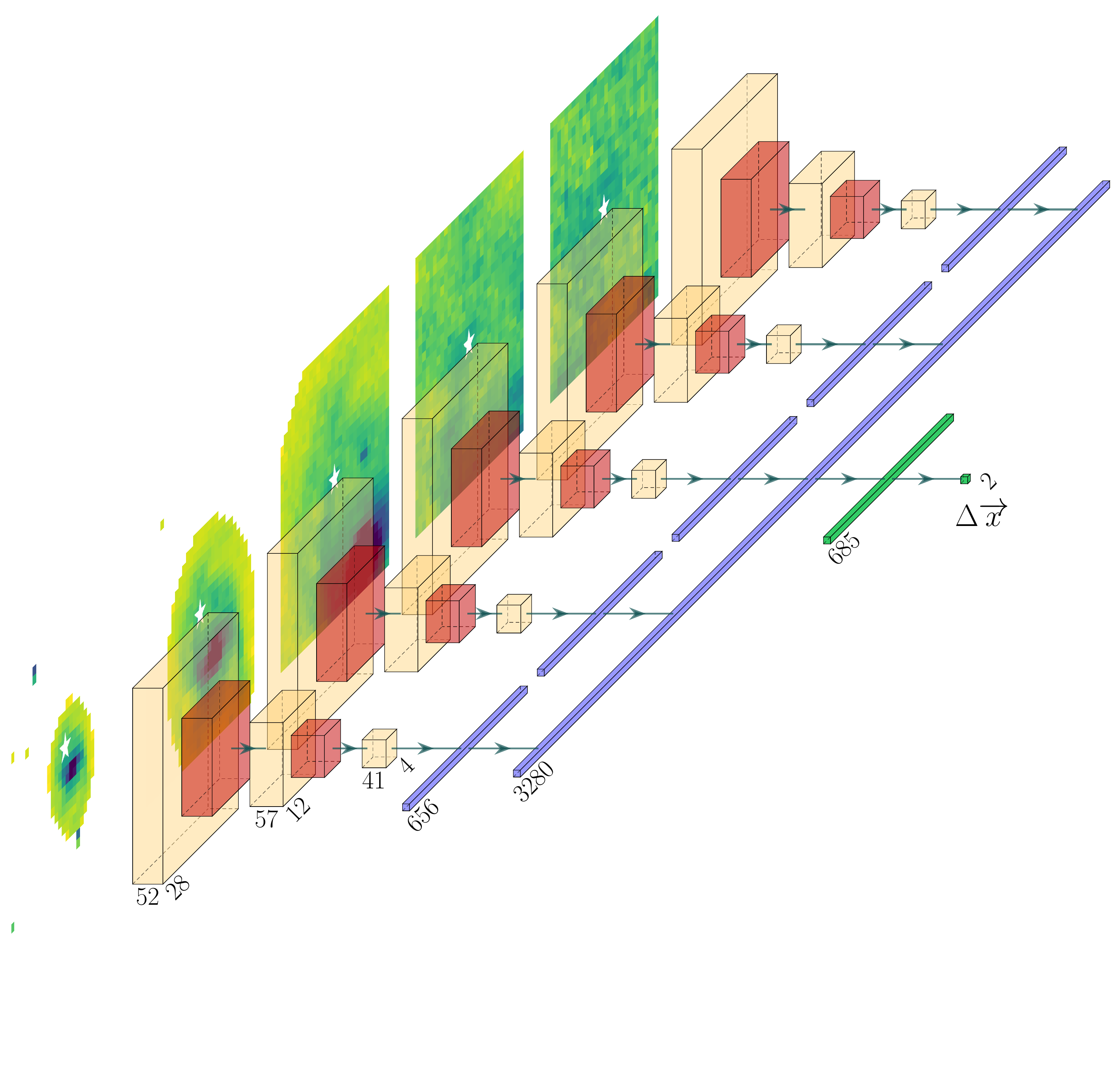}
    \put(3,8) {\large + 7 input sets}
    \put(3, 5) {\large (rotations and flips)}
    \put(84,44) {\large + 7 output vectors}
    \put(84,41) {\large (rotations and flips)}
    %\put(0,5.5) {\rotatebox{45}{\normalsize 30}}
    %\put(-1,8) {\rotatebox{90}{\normalsize 30}}
    \put(53,90) {\rotatebox{45}{\normalsize 30}}
    \put(59,90) {\rotatebox{90}{\normalsize 30}}
    \put(64.5,91.8) {\normalsize conv}
    \put(64.2,89.8) {\normalsize +relu}
    \put(67,85.8) {\normalsize max}
    \put(66.9,83.8) {\normalsize pool}
    \put(73,85) {\normalsize conv}
    \put(72.7,83) {\normalsize +relu}
    \put(75.1,73.1) {\normalsize max}
    \put(75,71.5) {\normalsize pool}
    \put(80.5,81.1) {\normalsize conv}
    \put(80.2,79.5) {\normalsize +relu}
    \put(88.5,83.2) {\normalsize flatten}
    \put(95,80.5) {\normalsize concat}
    \put(83,61.6) {\normalsize dense}
    \put(82.7,59.6) {\normalsize +tanh}
    \put(83.5,54.5) {\normalsize dense}
    \end{overpic}
  \caption{DELIGHT architecture diagram. The input of the network are five $30\times 30$ pixel images centered around the position of the candidate with different resolutions and fields of view (see Figure~\ref{fig:SN2004eh}). This set of images is rotated and flipped to generate seven more sets used as parallel inputs to the network. Each image within each set is processed by a combination of convolutional layers with $3 \times 3$ kernel size, no padding, and ReLU activation functions (conv+relu), shown as yellow rectangular cuboids; and $2 \times 2$ max-pooling layers (max pool), shown as red rectangular cuboids. At this stage we use the same weights and biases for every image given that the problem is approximately scale-free. Each output is then flattened into 656 neuron vectors (purple rectangular cuboids), and the five flattened outputs are concatenated into a layer of 3280 neurons (purple rectangular cuboid). This layer is fully connected with a layer of 685 neurons using a $\tanh$ activation function (green cuboid), which is then fully connected to the two neuron output layer (small green cuboid) without activation function representing the predicted host center in pixels. Each of the eight sets of five images produces an output vector, which can be rotated/flipped to produce a cloud of eight predicted host centers per candidate.}
  \label{fig:arch}
\end{figure*}
\setlength{\abovecaptionskip}{2pt plus 3pt minus 2pt}

We trained the network using Tensorflow 2.9.1 and Keras using a mean squared error loss function and Adam optimization \citep{kingma2014method}. The format of the input images, the architecture of the network, and the training strategy were chosen using $k$-fold cross-validation in the training set with $k=5$, after performing a train-test split using \ntest samples as test set. 

For the input images, we tried varying the number of levels and how we masked the images, either using a SExtractor mask, a median absolute deviation threshold mask, or no mask. For the architecture, we tried using either shared or independent convolutional layers for all the levels; using rotations and flips or not using them; using a dropout layer with different dropout factors; varying the number of channels and size of the convolution kernels; varying the number and size of the fully connected layers; and changing the type of activation functions. For the training strategy, we tried different learning rates and batch sizes.

After an initial exploration we chose to share convolutional layers among levels, use rotations and flips, $3\times3$ convolution kernels, and Rectified Linear Unit (ReLU), and $\tanh$ activation functions for the convolutional and fully connected layers, respectively. Although our analysis was limited to galaxies within one arcmin from the transient, we found that including host galaxies further away from the transient (in the corners of a square of side two arcmin centered around the transient) during training helped to improve the validation loss at large angular separations by more than a factor of two. 

We then performed a detailed hyper-parameter exploration. We assumed 4, 5 or 6 levels for the input images; the three masking strategies described before; a log-uniform distribution for the learning rate, between $10^{-4}$ and $10^{-1.5}$; a log-uniform distribution for the batch size, between 16 and 64; a log-uniform distribution for the number of convolution channels, between 16 and 64 for each convolutional layer; a log-uniform distribution for the number of neurons in the fully connected layer, between 128 and 512; and a uniform distribution for the dropout regularization factor, between 0 and 0.4. We explored 200 parameter combinations in Ray Tune\footnote{\url{https://docs.ray.io/en/latest/tune/index.html}} using the ASHA scheduler \citep{ASHA}, with up to 50 epochs during training and a grace period of 20 epochs, where a random fold is chosen at each iteration. 

We then selected the three best combinations of parameters and used the one with the best average validation loss between the five folds. Upon examination of the errors of the largest angular diameter galaxies, we decided to balance each batch during training by oversampling the galaxies from the nearby sample and repeating the hyperparameter search. In this second hyper parameter search we fixed the number of levels to 5 (the best number of layers according to the previous optimization) and used a fixed SExtractor masking strategy. We used the same distributions of parameters previously described, but we changed the range of possible number of neurons in the dense layer to be between 256 and 2048. We again selected the three best combinations of parameters and used the one with the best average loss between the five folds.

Finally, we re-trained the network with the previous hyper-parameters, but using the entire training set. The predicted  positions in the test set were then used for all the comparisons with other host galaxy identification methods, and the dispersions in the metrics of the five folds were used as metric errors. The best model training batch size was 40, and its learning rate, 0.0014. The resulting best architecture is summarized in Table~\ref{tab:arch} and Figure~\ref{fig:arch}. The typical prediction time per candidate for a set of multi-resolution images was 60 milliseconds using a CPU.

\section{Results} \label{sec:results}

\subsection{Host association error} \label{sec:errors}

Using the best model described in Section \ref{sec:arch}, we evaluate the results in the test set, using the average of the eight derotated and deflipped output displacement vectors as prediction. The resulting root mean squared deviation between the labeled and predicted positions was $1.836\pm0.051$\arcsec, the mean deviation was $0.783\pm0.009$\arcsec, the median deviation was $0.468\pm0.008$\arcsec, and the mode was $0.427\pm0.051$\arcsec.

\begin{figure}[ht!]
    \includegraphics[width=1\linewidth]{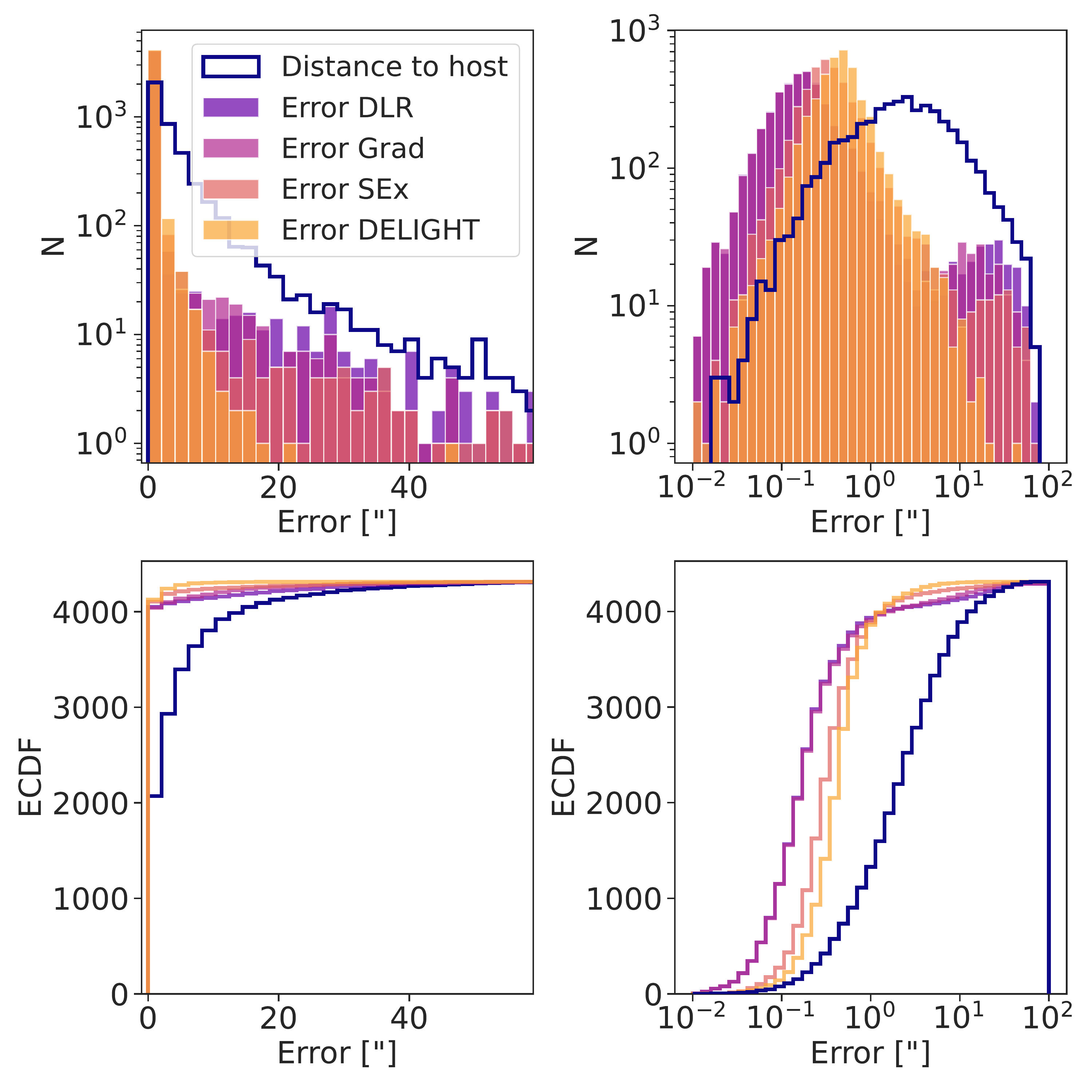}
  \caption{Distribution of errors between the predicted and labeled positions of the hosts using different methods (see Section~\ref{sec:methods}) compared to the distribution of host angular separations from the position of the transient. We use both linear and logarithmic scales in the errors (left and right, respectively), and a regular and cumulative histogram (top and bottom, respectively).}
  \label{fig:distances_comp}
\end{figure}

In Figure~\ref{fig:distances_comp} we show a comparison between the distributions of errors using DELIGHT, SEx, Grad, and DLR methods, compared to the distribution of angular separations between the transient and their host. We use only those transients where a comparison could be made, since in a fraction of the cases some of the methods do not yield a predicted host galaxy position (2.7\% and 9.5\% of the cases have no predicted position in DLR and Grad, respectively). In the left and right panels of this figure, we show errors in linear and logarithmic scales, respectively; and in the top and bottom panels, histograms and cumulative distributions of errors, respectively. It can be observed that DELIGHT leads to fewer catastrophic errors in general. DELIGHT is also the least precise method at sub-arcsecond scales.

\begin{figure}[ht!]
  \centering
    \includegraphics[width=\linewidth]{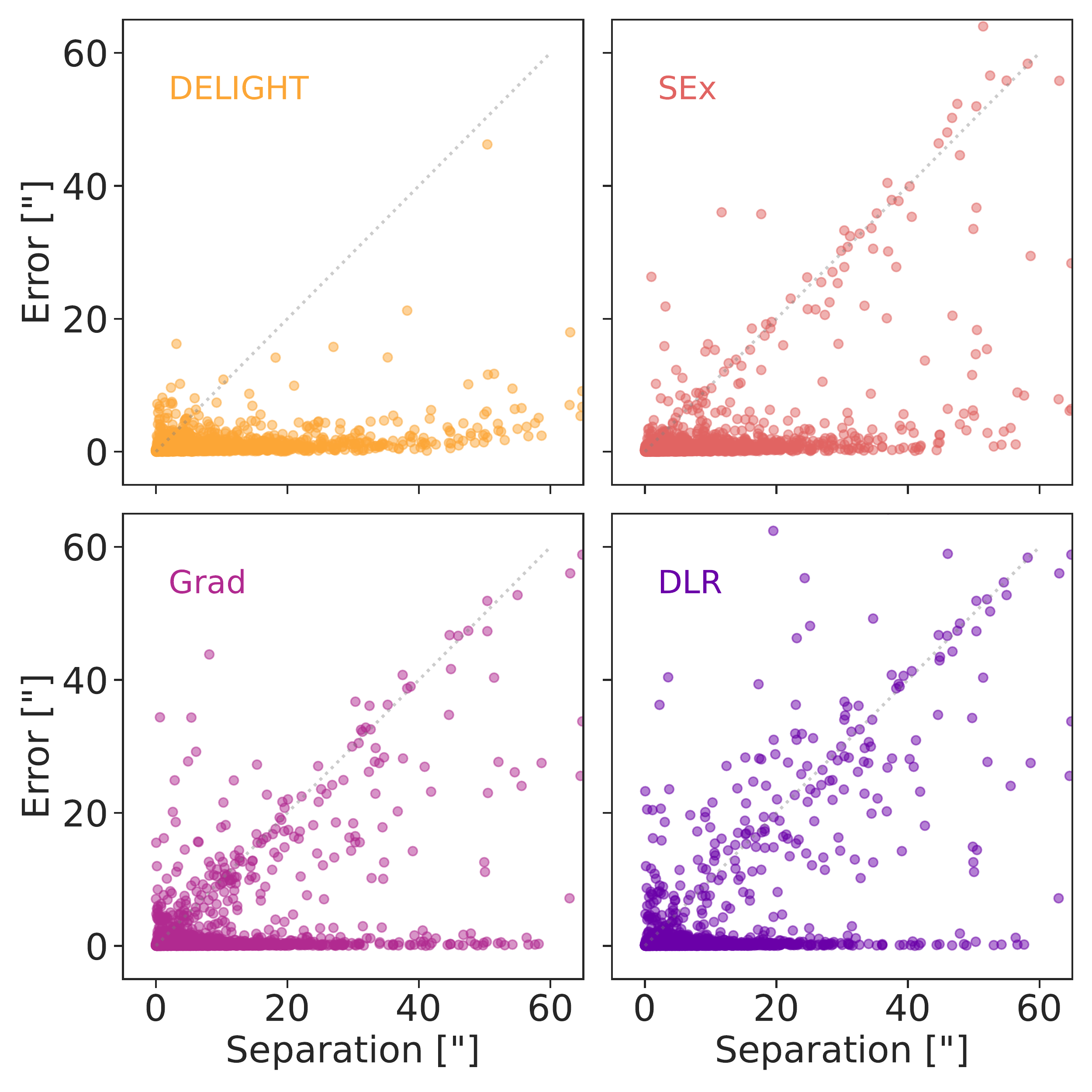}
  \caption{Prediction error vs angular separation distance between the transient candidate and the host galaxy center for the different methods tested in this work. Catastrophic errors are clustered near the identity line (gray, dotted lines). These errors correspond to cases where a large nearby host galaxy is confused with a smaller, distant galaxy which is closer in projection. }
  \label{fig:error_comp}
\end{figure}

In order to understand when the different methods lead to catastrophic errors, we study the distribution of errors as a function of angular separation to the host. In Figure~\ref{fig:error_comp} we show scatter plots of the errors vs. the angular separation to the host in the cases where the methods used in this analysis can be compared. It can be seen that most other methods show a small fraction of cases near the identity line, which corresponds to cases where the predicted position is very close to the transient position, but the true host is much further away in the plane of the sky, i.e. the host is confused with a smaller angular size, probably more distant in the line of sight galaxy. We do not see this behavior in DELIGHT, suggesting that the proposed method is able to accurately predict the host position and leads to fewer catastrophic errors even in large nearby galaxies with complex geometries. 

\begin{figure}[ht!]
  \centering
    \includegraphics[width=\linewidth]{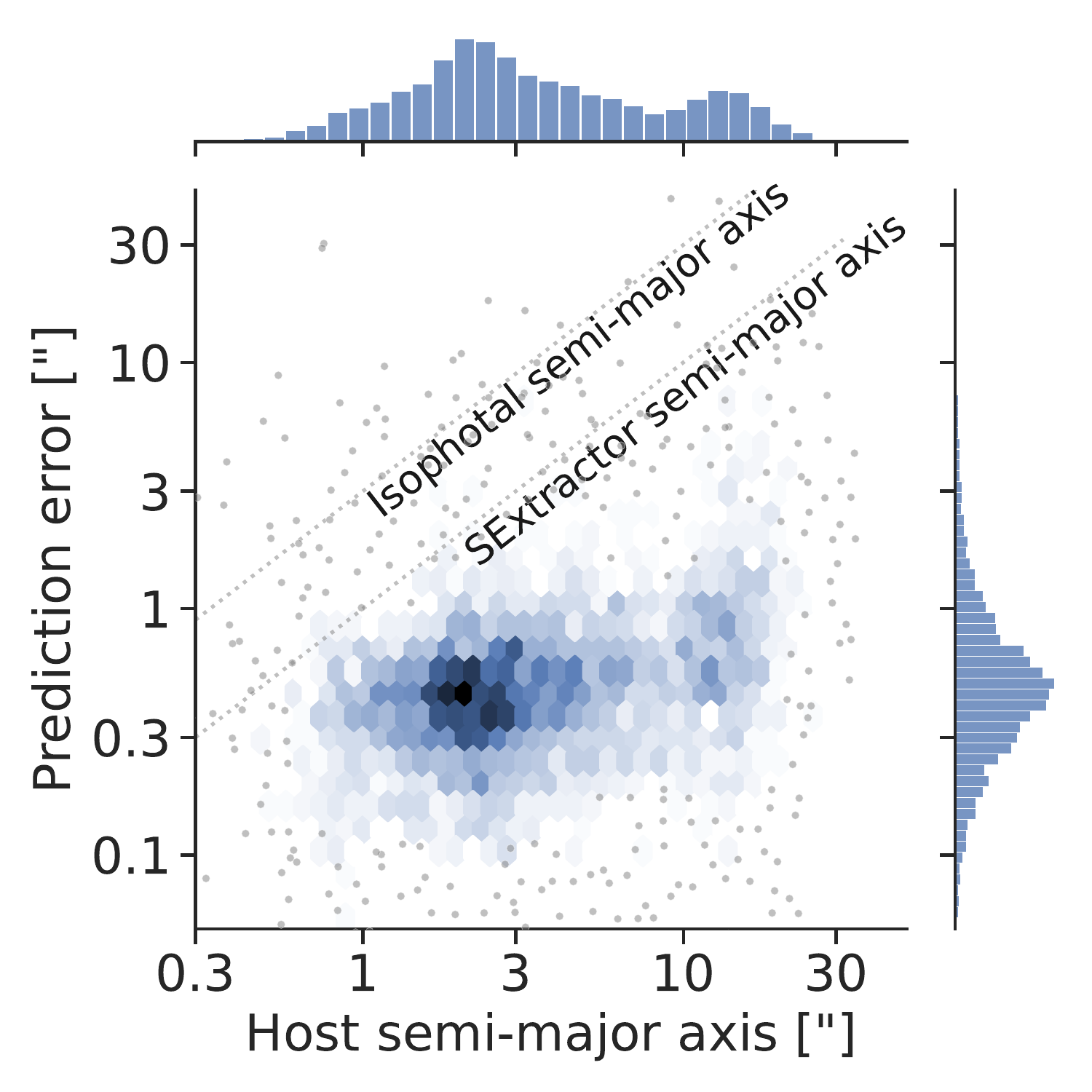}
  \caption{Distribution of the DELIGHT prediction errors and SExtractor inferred host semi-major axes using logarithmic bins. Dotted lines indicate the region where errors would equal the SExtractor semi-major axis and the inferred isophotal semi-major axis (three times the SExtractor semi-major axis).}
  \label{fig:2dhist}
\end{figure}

In Figure~\ref{fig:2dhist} we show the two-dimensional distribution of prediction errors vs host semi-major axes in logarithmic scale. We present all the test sample, without excluding examples where the comparison methods fail to predict a position as in Figures~\ref{fig:distances_comp} and \ref{fig:error_comp}. We use hexagonal bins and show those bins containing more than one example, and the rest of the data as a cloud of points. We also include the marginalized distributions in the top and right axes. In this figure, we also indicate the lines where the errors equal the host semi-major axis reported by SExtractor, and where the errors equal the inferred isophotal semi-major axis, a value that is in better agreement with what a person would visually estimate to be the size of the host (see Appendix~\ref{sec:appendix_diameter}). It can be concluded that in galaxies smaller than approximately 10\arcsec, the errors are typically below 1\arcsec\ in absolute value, and in larger galaxies the distribution of errors has a significant tail extending up to a few arcseconds in absolute value. 

Finally, in Appendix~\ref{sec:appendix_examples} we show ten random examples in the ALeRCE (Figure~\ref{fig:examples_random}) and Nearby samples (Figure~\ref{fig:examples_random_nearby}) to visualize how the proposed method typically performs, as well as the ten worst normalized error cases (Figure~\ref{fig:examples_worst}) and the ten worst normalized standard deviation cases of the DELIGHT predictions (Figure~\ref{fig:examples_spread}) in order to give a sense of when this method fails.

\subsection{Redshift association error} \label{sec:redshifts}

\begin{deluxetable}{ccc}[ht!]
\tablecaption{Number of spectroscopic and photometric redshift cross-matches in SDSS DR17 using different host identification methods as input positions.}
\label{tab:xmatches}
\tablehead{
\colhead{Method} & \colhead{\# spec-z matches} & \colhead{\# photo-z matches}}
\startdata
True position & 1079 & 3031 \\
DELIGHT & 1063 & 3010 \\
SEx & 1041 & 2999 \\
DLR & 988 & 2889 \\
Grad & 940 & 2752 \\
1-NN & 736 & 2655 \\
\enddata
\tablecomments{DELIGHT is the method that yields more spectroscopic and photometric redshift cross-matches other than using the true host position.}
\end{deluxetable}

An additional test that was conducted was to study the redshift association error that results when using the predicted host positions and cross matching them with a catalog of host galaxy redshifts. This is important because it is generally more important to recover the redshift of the host rather than its exact position. We used the SDSS DR17 Object CrossID service\footnote{\url{https://skyserver.sdss.org/dr17/en/tools/crossid/crossid.aspx}} to obtain the closest galaxy source with spectroscopic or photometric redshift within 0.5 arcmin from the predicted test set positions. The number of matches using different methods for host determination is shown in Table~\ref{tab:xmatches}. Note that DELIGHT is the method that yields more spectroscopic or photometric redshift matches other than using the true host position. 

\begin{figure}[ht!]
  \centering
    \includegraphics[width=\linewidth]{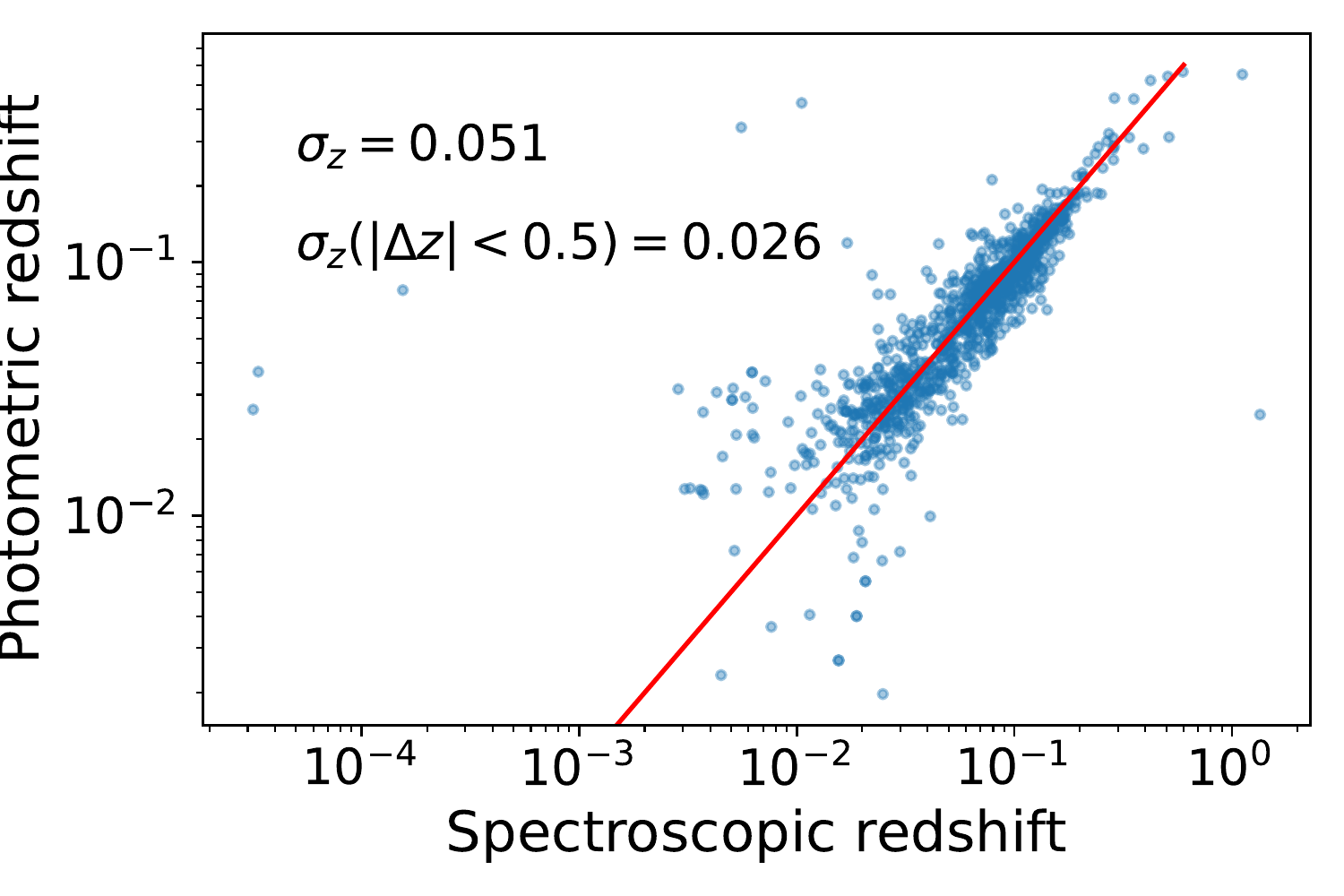}
  \caption{Comparison between spectroscopic and photometric redshifts in the test set crossmatched sample. The photometric error dispersion considering all the samples is 0.051, and the dispersion after removing redshift errors with absolute value larger than 0.5 is 0.026.}
  \label{fig:specvsphot}
\end{figure}

\begin{figure}[ht!]
  \centering
    \includegraphics[width=\linewidth]{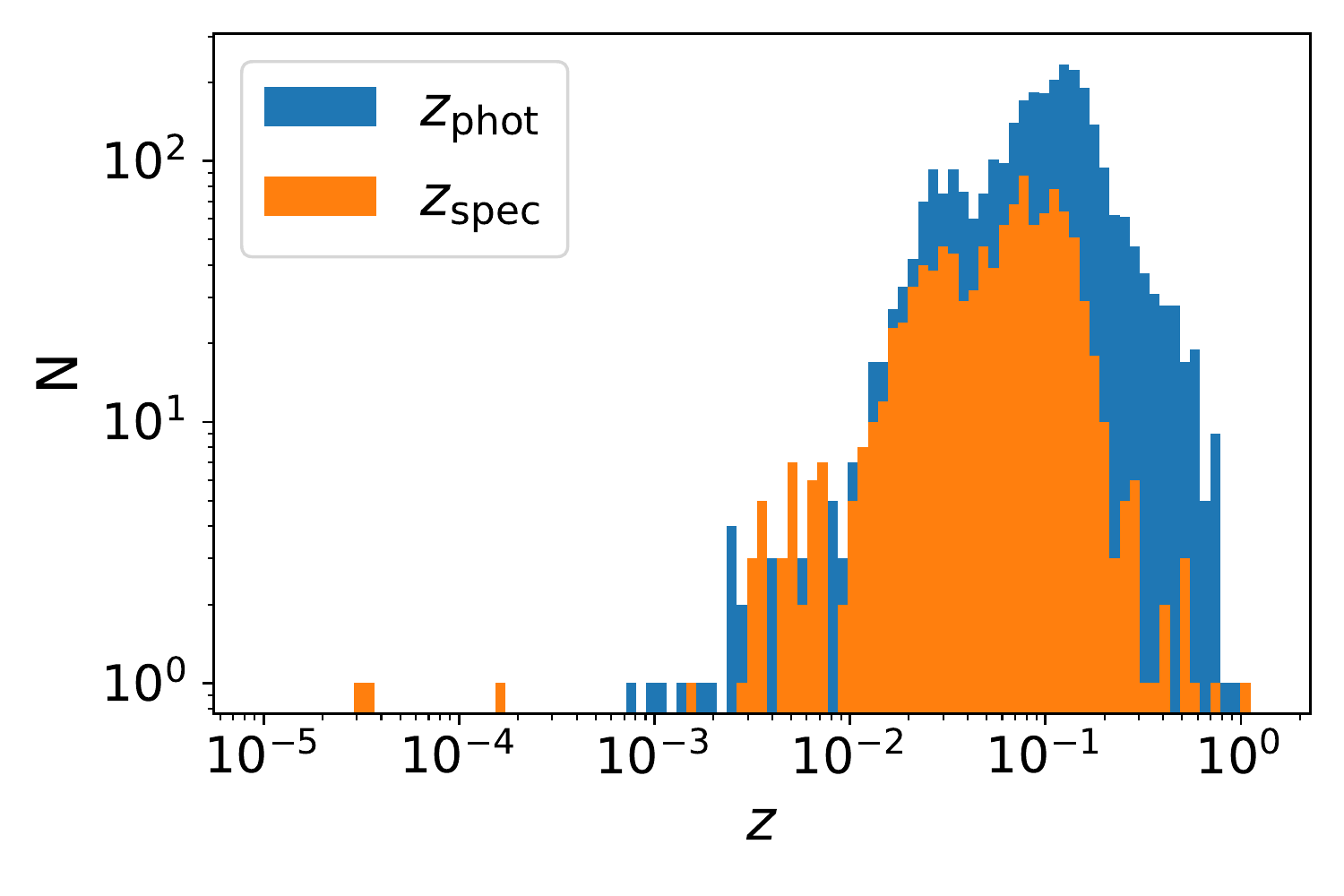}
  \caption{Histogram of test set spectroscopic and photometric redshifts obtained after cross-matching the labeled host positions with SDSS DR17 using a search radius of 0.5 arcmin. }
  \label{fig:zhosts}
\end{figure}

A comparison between photometric and spectroscopic errors suggests that the photometric redshift error is typically of the order of 0.051 (or 0.026 if we reject 2 outlier samples with $|\Delta z| > 0.5$), as shown in Figure~\ref{fig:specvsphot}. The distribution of all the cross-matched spectroscopic and photometric redshifts in the test set is also shown in Figure~\ref{fig:zhosts}. A bimodality is observed in the distribution that originates from the combination of both the Nearby and ALeRCE samples.

\begin{figure*}[ht!]
  \centering
  \includegraphics[width=\linewidth]{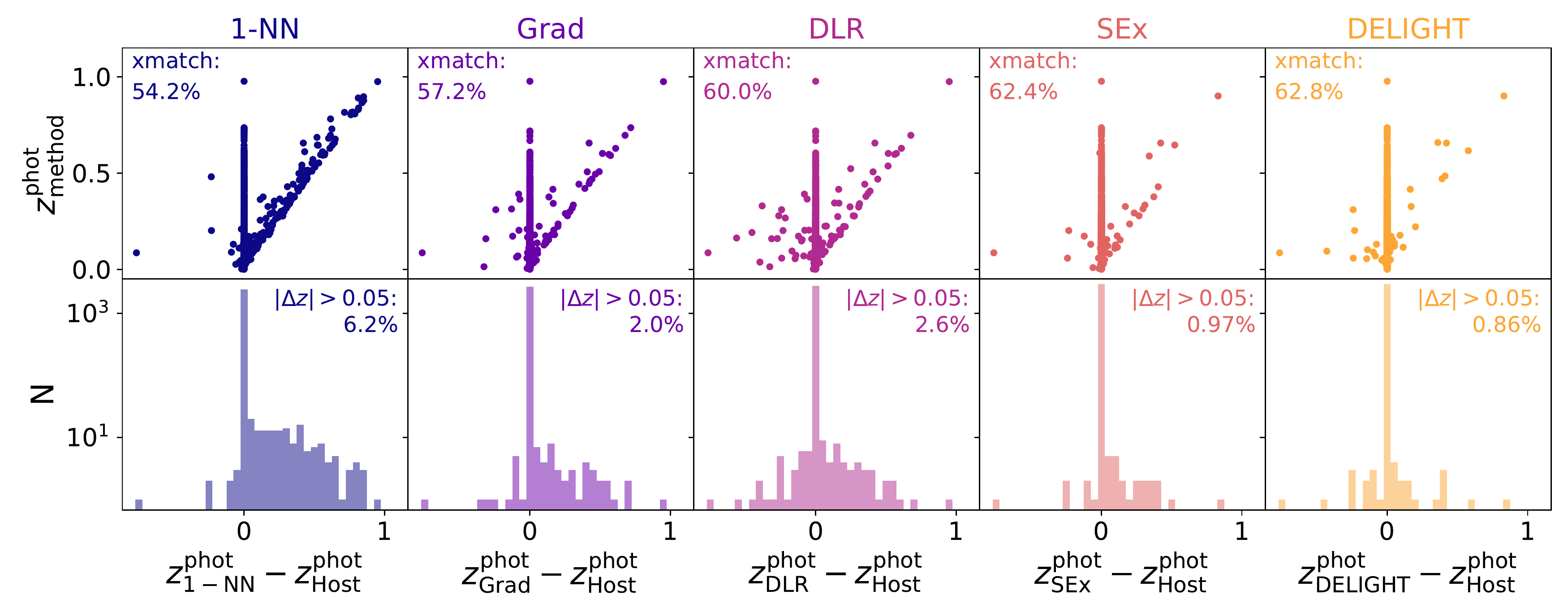}
  \caption{Crossmatched photometric redshift differences. \textit{Top row}: scatter plots of the redshifts determined using the method's predicted position vs. its difference with the redshift determined using the true host position, indicating the fraction of crossmatches with the given method (a proxy for completeness). \textit{Bottom row}: histograms of the redshift errors, indicating the fraction of errors whose absolute value is larger than 0.05, a proxy for contamination. }
  \label{fig:z_comp}
\end{figure*}

Finally, in order to compare the accuracy in the redshift determination methods, in Figure~\ref{fig:z_comp} we show the distribution of redshift errors using different host identification methods and cross-matching with SDSS DR17. The top row shows the crossmatched photometric redshift inferred from the position predicted by the different methods using SDSS DR17 vs. its difference with the redshift inferred from the labeled host position using SDSS DR17, indicating the crossmatch completeness at the top left corner. The bottom row shows the histogram of differences in logarithmic scale with bins of $\Delta z=0.05$, indicating the fraction of redshift differences whose absolute value is greater than 0.05 (larger than the dispersion found in Figure~\ref{fig:specvsphot}), a proxy for contamination. Note that in the top row most incorrect redshifts lie close to the identity line and with positive redshift differences, which is what would be expected if large angular separation, nearby hosts are being confused with small angular separation, more distant, unrelated hosts.

It can be seen that DELIGHT outperforms all the methods in both completeness and contamination in the test set. A summary of the completeness and contamination of the different methods using different redshift types and samples is shown in Table~\ref{tab:summary}. This shows that DELIGHT is the most complete method for all redshift types and samples, and the least contaminated method in almost all redshift types and samples. Our method is only outperformed by SEx in contamination for the ALeRCE sample using photometric redshifts.

\begin{deluxetable*}{ccccccccccccc}[ht!]
\tablecaption{Completeness and contamination when recovering the host spectroscopic and photometric redshifts using different methods and samples.}
\label{tab:summary}
\tablehead{
 & \multicolumn{6}{c}{Spectroscopic redshifts} & \multicolumn{6}{c}{Photometric redshifts} \\
 & \multicolumn{3}{c}{Completeness} & \multicolumn{3}{c}{Contamination}  & \multicolumn{3}{c}{Completeness} & \multicolumn{3}{c}{Contamination} \\ 
 \colhead{Method}  & \colhead{All} & \colhead{ALeRCE} & \colhead{Nearby} & 
 \colhead{All} & \colhead{ALeRCE} & \colhead{Nearby} &
 \colhead{All} & \colhead{ALeRCE} & \colhead{Nearby} &
 \colhead{All} & \colhead{ALeRCE} & \colhead{Nearby} }
\startdata
True position & 22.6 & 20.4 & 32.6 & & & & 63.4 & 65.9 & 51.1 \\
DELIGHT & \bf{22.2} & \bf{20.2} & \bf{31.7} & 0 & 0 & 0 & \bf{62.8} & \bf{65.6} & \bf{49.9} & \bf{0.86} & 0.93 & \bf{0.48} \\
SEx & 21.8 & 20.1 & 29.7 & 0 & 0 & 0 & 62.4 & \bf{65.6} & 47.5 & 0.97 & \bf{0.66} & 3.0 \\
DLR & 20.7 & 19.7 & 25.4 & 0 & 0 & 0 & 60.0 & 63.5 & 43.7 & 2.6 & 1.9 & 7.1 \\
Grad & 19.6 & 19.1 & 22.0 & 0 & 0 & 0 & 57.2 & 61.1 & 38.6 & 2.0 & 1.2 & 8.1 \\
1-NN & 15.4 & 16.8 & 8.4 & 0.14 & 0 & 1.4 & 54.2 & 60.9 & 22.9 & 6.2 & 3.4 & 40.8
\enddata
\tablecomments{We assume that redshifts that differ by more than 0.05 from the one obtained using the true position are equivalent to a misclassification. We use bold letters for the best metric in each category other than using the true position. Note that DELIGHT outperformed all other methods in completeness and purity, except in purity for the ALeRCE photometric redshift sample using SEx.}
\end{deluxetable*}

\section{Conclusions} \label{sec:conclusions}

We have introduced DELIGHT, a novel host galaxy identification method that uses multi-resolution archival images centered at the position of extragalactic transient candidates using a convolutional neural network in order to predict the position of the host. We have compared this method to other state-of-the-art host galaxy identification methods and found that it outperforms all of them in completeness and contamination when recovering host galaxy redshifts using the predicted position and cross-matching with SDSS DR17, especially with large angular diameter, nearby galaxies. Although this method is the least precise in identifying the host galaxy center at sub-arcsecond scales, it is the method with fewer catastrophic association errors, which explains the better inferred redshifts. 

To the best of our knowledge, this simple but competitive model serves as a first example of the application of machine learning techniques for the problem of direct and automatic transient host identification based solely on images, yielding the correct host position even if the host galaxy has not been catalogued. The model is currently being used by the ALeRCE broker team to automatically annotate the host galaxies of transient candidates from the ZTF public stream submitted to TNS using archival $r$-band PanSTARRS images and simple cuts to ensure that the predicted positions are matched to the correct sources in external catalogs, and visually inspecting these annotations in order to ensure that no mismatches are sent to the community. We foresee developing more advanced versions of this model in the future, including returning multiple host candidates in the cases where the identification is degenerate. 

Our method is very competitive in terms of speed if multi-resolution images are available, taking about 60 milliseconds per transient to determine the host position using a CPU after the image is loaded in memory (c.f. 1.1 s in Grad, the other image-based method). This is fast enough to be applied in real-time to the entire alert streams of ZTF and LSST with current broker computing capabilities. Additionally, the input images have a size of only 3.5\% the size of the input images with the same scale and a constant resolution (32 kB vs 920 kB, respectively), which is smaller than the LSST alert packets (32 kB vs. 80 kB), although larger in size to the postage-stamp images contained in these alerts (about 20 kB). This could be decreased even further by removing the central parts of the larger field-of-view images in the multi-resolution set (they are redundant with the next, higher resolution level),  by changing the total field-of-view and number of levels of the set, or by using different data compression techniques. 

While the ZTF postage-stamp images are 63\arcsec $\times$ 63\arcsec\ (63 pix $\times$ 63 pix), the LSST postage-stamp images are expected to be only 6\arcsec $\times$ 6\arcsec\ (30 pix $\times$ 30 pix) in size, which will make the problem of automatically classifying postage-stamp images very difficult (c.f. \citealt{2021AJ....162..231C}). Thus, we highlight the real possibility of using some form of large field-of-view multi-resolution postage-image stamps for both the problems of rapid postage-stamp image classification and host association in the Vera C. Rubin Observatory alert stream. This would require that the multi-resolution images are generated while the raw LSST images are in memory, before the alert packets are distributed to the different alert brokers. Alternatively, a special service for obtaining multi-resolution images for LSST transient candidates could aid with the host galaxy identification, especially in large angular separation hosts as a post-processing step close to real-time.

A working version of DELIGHT is available via the Python Package Index (PyPI) in \url{https://pypi.org/project/astro-delight/} and can be installed using the command line \verb+pip install astro-delight+. The source code and example notebooks are available in GitHub in \url{https://github.com/fforster/delight}.

\begin{acknowledgments}

The authors acknowledge support from
the Chilean Ministry of Economy, Development, and Tourism's Millennium Science Initiative through grant ICN12\textunderscore 12009, awarded to the Millennium Institute of Astrophysics (AMA, FEB, FF, GP, IR, JCO, LHG, MC, PE, PH, PSS, RD), 
and from the National Agency for Research and Development (ANID) grants: 
BASAL Center of Mathematical Modelling Grant PAI AFB-170001 (FF, AMA, IR), BASAL project FB210003 (MC, FEB),
FONDECYT Regular 1200710 (FF), FONDECYT REGULAR 1200495 (FEB, JCO), FONDECYT REGULAR 1211374 (PH),
FONDECYT Postdoctorado 3220449 (RD),
National Doctorate Grant 21221393 (SCC), ANID-PFCHA/Doctorado-Nacional/2020-21202606 (MR), ANIDPFCHA/Doctorado Nacional/2019-2119188 (DG), and infrastructure funds QUIMAL140003 and QUIMAL190012.
%
%\end{acknowledgments}
%\begin{acknowledgments}
We acknowledge support from REUNA Chile.
We are thankful to the La Serena School of Data Science \citep{2021IAUS..367..458B}, where the first proof-of-concept version of this method was implemented.
We are thankful for the support provided by Fintual.
A.G. is supported by the National Science Foundation Graduate Research Fellowship Program under Grant No.~DGE–1746047. A.G. further acknowledges funding from the Center for Astrophysical Surveys Fellowship at UIUC/NCSA and the Illinois Distinguished Fellowship. 
Based on observations by the Zwicky Transient Facility (ZTF, see full acknowledgments in \url{https://irsa.ipac.caltech.edu/data/ZTF/docs/releases/dr07/ztf_release_notes_dr07.pdf}), the Pan-STARRS1 Surveys (PS1, see full acknowledgments in \url{https://ipp.ifa.hawaii.edu/}), and the Sloan Digital Sky Survey (SDSS, see full acknowledgments in \url{https://www.sdss.org/collaboration/citing-sdss/}).
This research has made use of the NASA/IPAC Extragalactic Database (NED) and the SIMBAD database.
\end{acknowledgments}

\begin{appendix}

\counterwithin{figure}{section}

\section{Measuring the hosts semi-major axis using multi-resolution images} \label{sec:appendix_diameter}

\begin{figure*}
\centering
\vbox{
  \includegraphics[width=\linewidth]{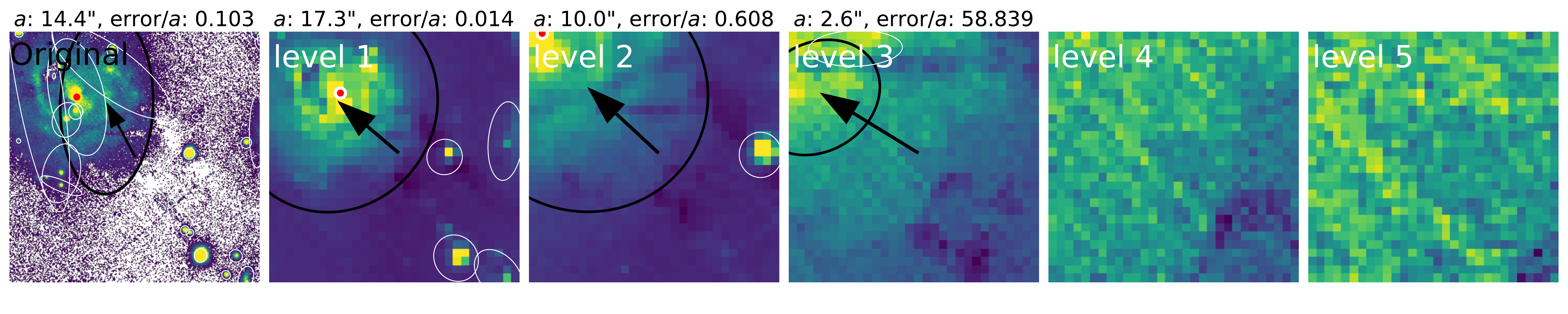}
    \includegraphics[width=\linewidth]{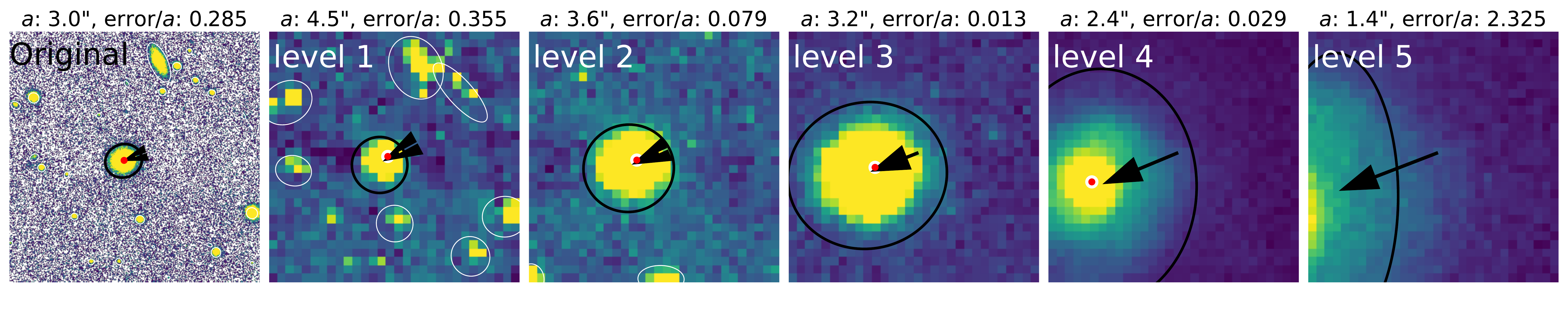}
    \includegraphics[width=\linewidth]{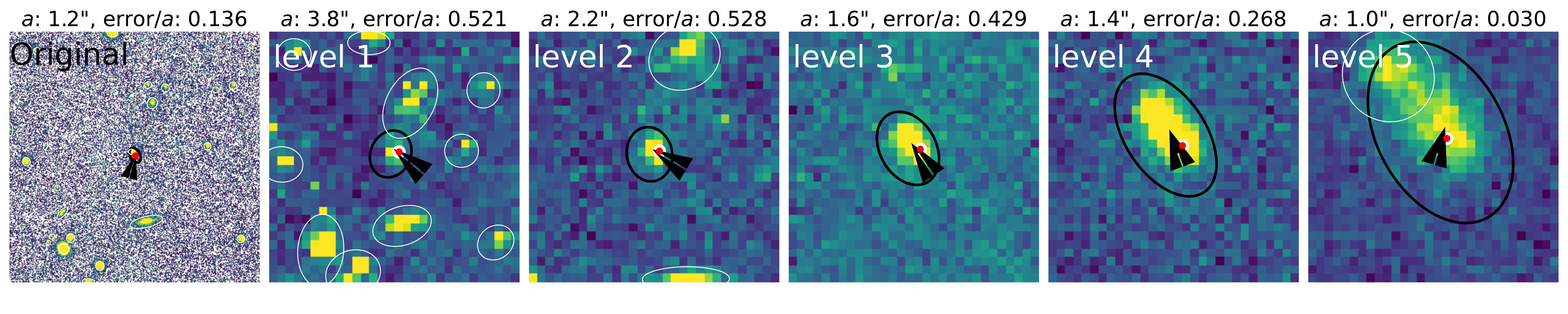}
  }
  \caption{Three examples of how the semi-major axis of the galaxies were determined. In the top row we show SN2009jy; in the middle row, ZTF19acbwhqc; and in the bottom row, ZTF19acbwmqd. We show the original 480 pix $\times$ 480 pix PanSTARRS $r$-band image in the first column, and the five central 30 pix $\times$ 30 pix multi-resolution images in the following columns. The SExtractor detected object ellipses are shown, with the ellipse in black having the closest normalized distance from its center to the labeled position. The black arrow points to the best ellipse center and the white and red dot marks the labeled host position. In the title, $a$ is the semi-major axis and error/$a$ is the distance between the object's inferred center and the labeled position divided by the semi-major axis. In the top row, the nearest normalized distance object detected in level 1 is coincident with the labeled position within 1.4\% of the inferred semi-major axis, so the associated semi-major axis, 17.3\arcsec, is accepted. In the middle row, the nearest normalized distance object detected in level 3 is coincident with the labeled position within 1.3\% of the inferred semi-major axis, so the associated semi-major axis, 3.2\arcsec, is accepted. In the bottom row, the nearest normalized distance object detected in level 5 is coincident with the labeled position within 3.0\% of the inferred semi-major axis, so the associated semi-major axis, 1.0\arcsec, is accepted. Note that the semi-major axis tends to be larger for lower resolution images, that are more sensitive to faint surface brightness regions of the host galaxy. Also note that SExtractor cannot represent the largest galaxy correctly in the original resolution because of all the small-scale structures visible.}
  \label{fig:examples_semimajor}
\end{figure*}

In order to determine the host galaxy semi-major axis we use SExtractor iteratively in a set of five multi-resolution unmasked images centered around the transient. These are constructed in a similar fashion as the method described in Section~\ref{sec:input}, but without any masking. The motivation is to avoid SExtractor from seeing small scale structure unless it is really needed. Starting from the largest field of view image, we use SExtractor to identify all the sources visible in the image and select the semi-major axis of the one closest to the true galaxy position in normalized elliptical distance. As long as the distance from the inferred position to the labeled host center is larger than 5\% the semi-major axis, we repeat the same measurement with a smaller field of view level. If the new normalized elliptical distance is smaller than in the previous level, we update the semi-major axis. The resulting semi-major axis is then used as a proxy for host size. The typical processing time per candidate is 150 milliseconds. Three examples of this algorithm are shown in Figure~\ref{fig:examples_semimajor}, where we show ellipses using the isophotal semi-major and semi-minor axes (three times the SExtractor semi-major and semi-minor axes, respectively). 

We tested this method via visual inspection of hundreds of cases and it gives good estimations of the semi-major axis for both small and large apparent size host galaxies, although with some underestimation of the semi-major axis for the largest galaxies (whose size is comparable or larger than the field of view of the image). We found that the SExtractor detected sources were coincident with the labeled host position within a radius of 5\% of the semi-major axis in 98\% of the cases. Only those cases were used to build Figure~\ref{fig:separations} to give a more robust representation of the distribution of semi-major axes. Upon visual inspection, we found that in the 2\% of cases where this condition was not met, 1) either the host center was outside the field of view, 2) the labeled position was significantly off the visually identified host, or 3) the host was not detectable with SExtractor; suggesting that this method could be used to further clean the sample. In fact, up to 11\% of the cases with no cross-matches and 31\% of the cases with incorrect redshifts described in Section~\ref{sec:redshifts} are found within this 2\% sample. We decide not to remove these examples to avoid biasing the results of this work to only those host galaxies that are well described with SExtractor ellipses.

\section{Examples of host galaxy predicted coordinates} \label{sec:appendix_examples}

\begin{figure*}
 \includegraphics[width=\linewidth]{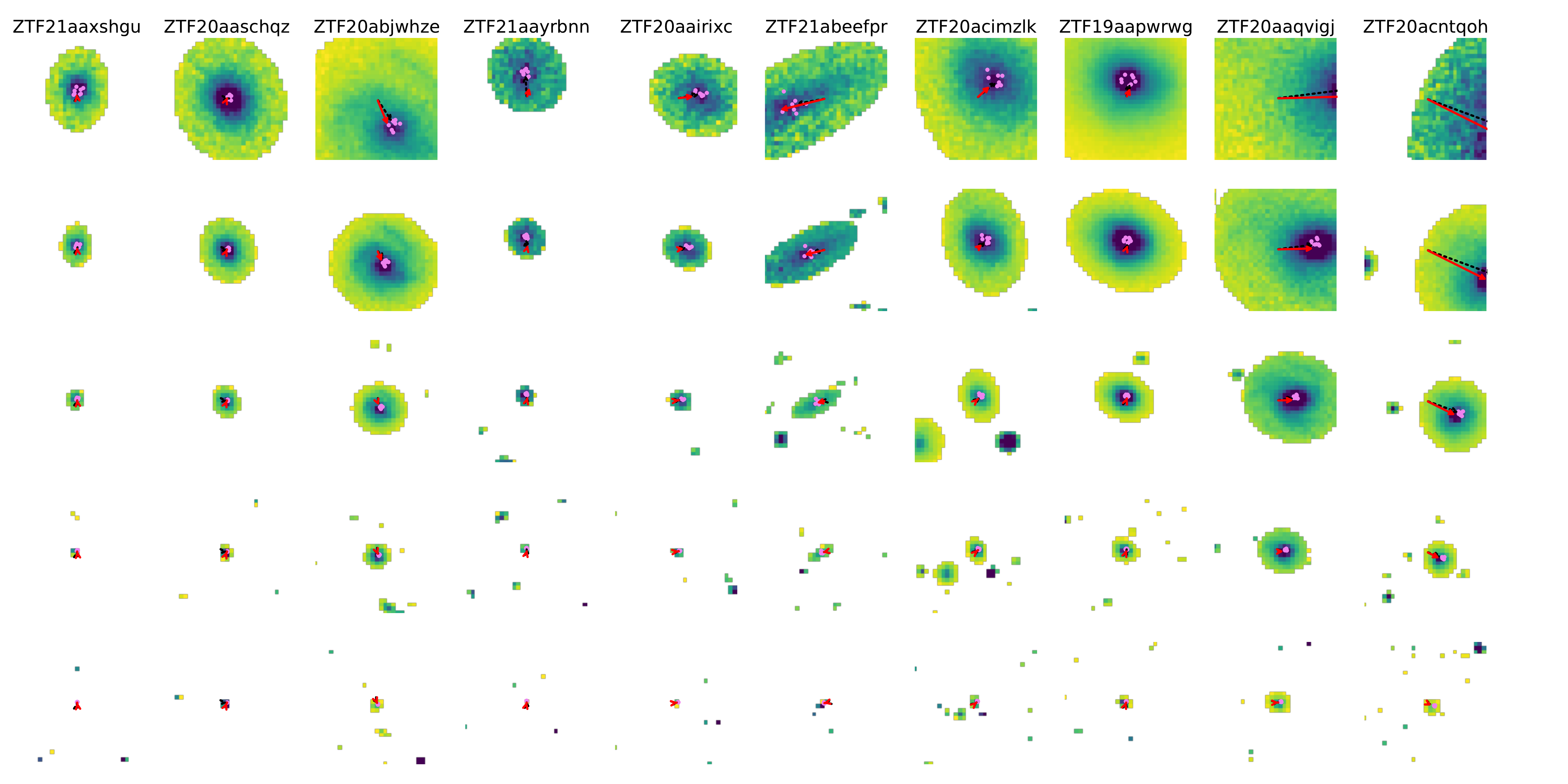}
  \caption{Comparison between the true host position and the DELIGHT predicted position in ten random examples from the ALeRCE sample. Each column shows the set of five multi-resolution images for a given galaxy, from smallest to largest field of view (highest to smallest resolution) from top to bottom. The red-solid arrows represent the true host position (obtained via visual crossmatching with catalog positions in NED, SIMBAD and SDSS DR16), the black-dotted arrows represent the mean DELIGHT predicted position, and the violet points represent the eight DELIGHT predicted positions for the eight rotations and flips. }
  \label{fig:examples_random}
\end{figure*}

\begin{figure*}[ht!]
 \includegraphics[width=\linewidth]{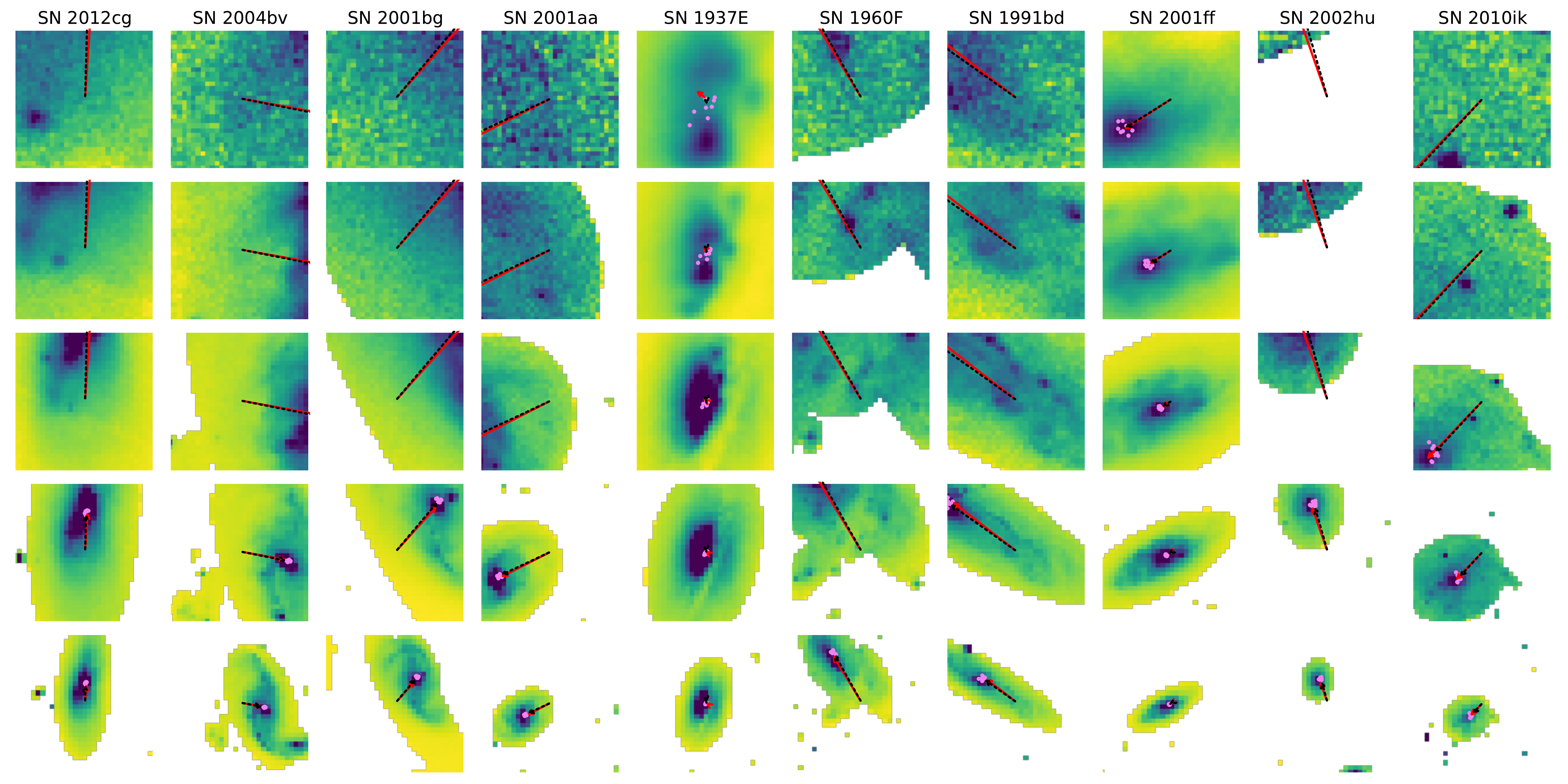}
  \caption{Same as Figure~\ref{fig:examples_random}, but for 10 random examples from the Nearby sample.}
  \label{fig:examples_random_nearby}
\end{figure*}

\begin{figure*}[ht!]
  \includegraphics[width=\linewidth]{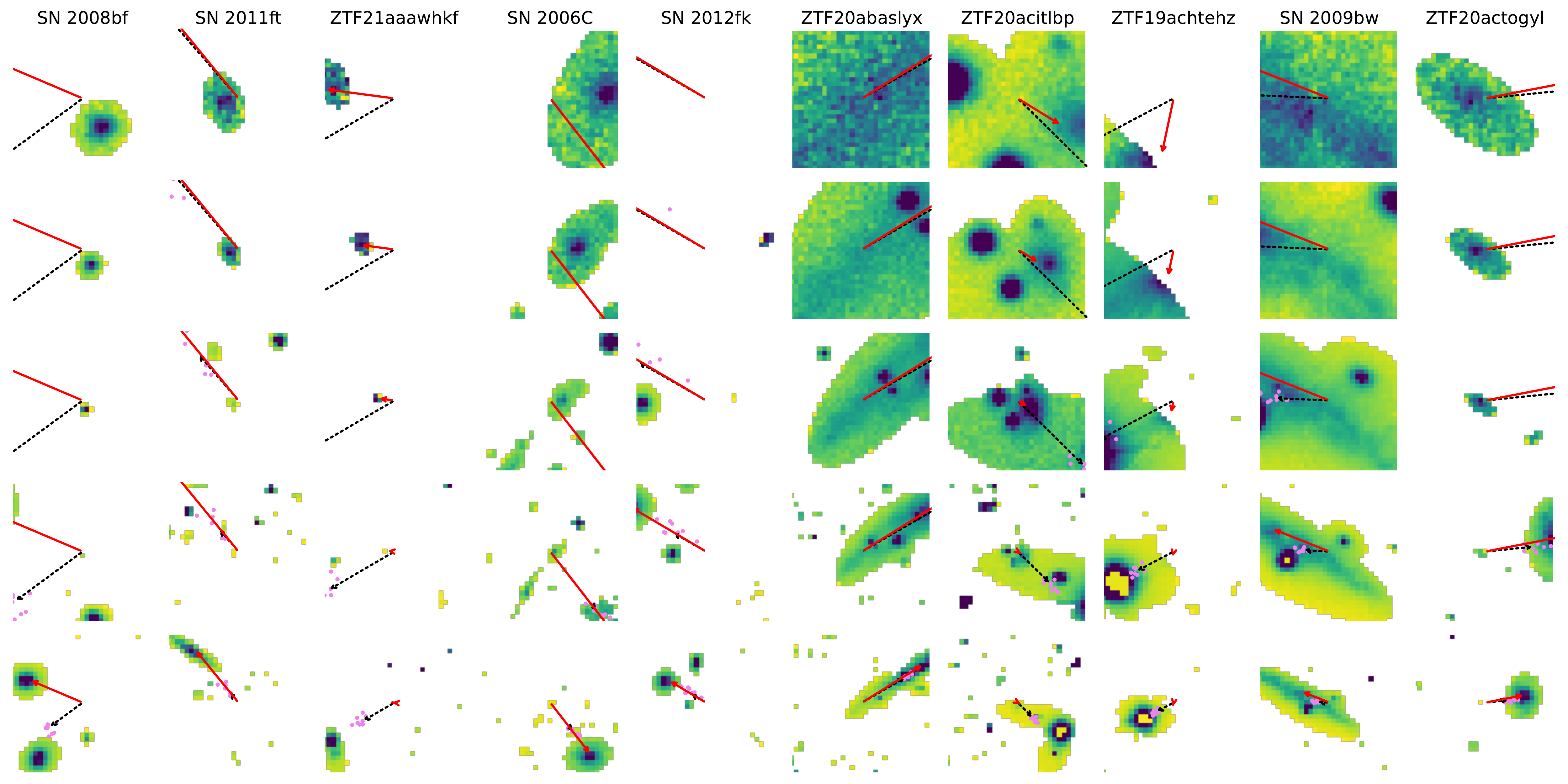}
  \caption{Same as Figure~\ref{fig:examples_random}, but for the ten examples with the largest errors normalized by angular separation among galaxies with an angular separation larger than 2.5\arcsec. Note that large normalized errors are seen in cases where there is a degeneracy between two or more galaxies as potential hosts, such as in \href{http://www.astrosurf.com/snweb2/2008/08bf/08bfHome.htm}{SN2008bf}, \href{http://observ.pereplet.ru/images/SN110830/MASTER110830.png}{SN20011ft}, \alercelink{ZTF21aaawhkf}, and \href{http://www.astrosurf.com/snaude/sn2012_8.htm\#2012fk}{SN2012fk}; when SExtractor masks HII regions in a given galaxy as different extended sources, such as in \href{http://www.astrosurf.com/snaude/sn2006_1.htm\#2006C}{SN2006C} and \alercelink{ZTF20actogyl}; and when there are bright stars close to or in the line of sight of the real host, such as in \alercelink{ZTF20abaslyx}, \alercelink{ZTF20acitlbp}, \alercelink{ZTF19achtehz}, and \href{https://www.rochesterastronomy.org/sn2009/sn2009bw.html}{SN2009bw}.}
  \label{fig:examples_worst}
\end{figure*}

\begin{figure*}
  \includegraphics[width=\linewidth]{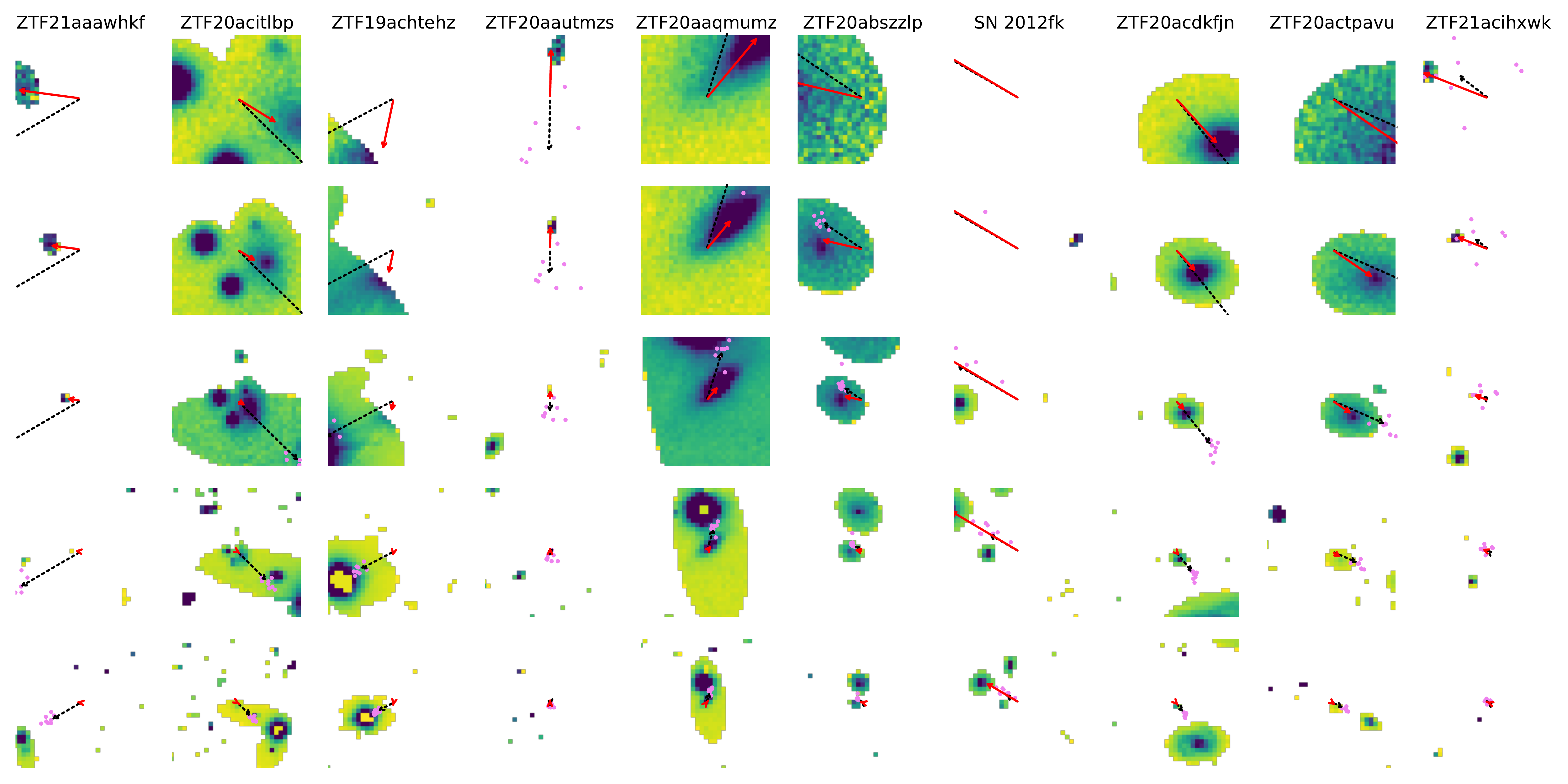}
  \caption{Same as Figure~\ref{fig:examples_random}, but for the ten examples with the largest standard deviation of the DELIGHT predictions normalized by angular separation, among galaxies with an angular separation larger than 2.5\arcsec . They tend to be similar cases as those found in Figure~\ref{fig:examples_worst}, with four of these largest normalized standard deviations also being among the top ten worst cases in terms of normalized errors.}
  \label{fig:examples_spread}
\end{figure*}

In this section we show individual predictions from DELIGHT compared to the true host position as visually selected from NED, SIMBAD and SDSS DR16 catalogs. First, in Figure~\ref{fig:examples_random} we show ten random examples from the ALeRCE sample of galaxies and in Figure~\ref{fig:examples_random_nearby}, ten random examples from the nearby sample of galaxies. Then, in Figure~\ref{fig:examples_worst} we show the ten worst examples in terms of errors normalized by angular separation, and in Figure~\ref{fig:examples_spread}, the ten worst examples in terms of standard deviation of the eight DELIGHT predicted positions normalized by angular separation. In the last two cases we restrict the sample to galaxies with a transient to host separation of at least 2.5\arcsec\ to avoid division by small numbers. In almost all cases DELIGHT is able to correctly identify the host position, as also shown in Figures ~\ref{fig:distances_comp} and \ref{fig:error_comp}, and for those cases in which it fails, the normalized standard deviation of the DELIGHT predictions can be used as a proxy for the normalized error.

\end{appendix}

\bibliography{DELIGHT}{}
\bibliographystyle{aasjournal}



\end{document}